\documentclass[twocolumn,trackchanges]{aastex63}
\usepackage{amsmath,amssymb,gensymb,times,graphicx,morefloats,color}
\usepackage{natbib,hyperref}
\usepackage[outdir=./]{epstopdf}

\newcommand\eg{\emph{e.g.}}

\newcommand\msun{\ensuremath{\text{M}_\odot}}

\newcommand{\mps}{m\,s$^{-1}$}

\newcommand{\vsini}{\ensuremath{v\sin{i_*}}}

\newcommand{\logg}{$\log{g}$ }

\newcommand{\um}{$\mu$m}
\newcommand{\fbol}{$F_{\mathrm{bol}}$}

\newcommand{\teff}{\ensuremath{T_{\text{eff}}}}
\newcommand{\gaia}{\textit{Gaia}}
\newcommand{\rprstar}{R_P/R_*}
\newcommand{\mearth}{\ensuremath{M_\oplus}}

\newcommand{\rearth}{$R_\Earth$}

\newcommand{\kms}{\ensuremath{\rm km~s^{-1}}}

\def\kepler{\emph{Kepler}}
\def\spitzer{\emph{Spitzer}}
\def\tess{\emph{TESS}}
\def\TESS{\emph{TESS}}

\def\hst{\emph{HST}}
\def\jwst{\emph{JWST}}
\def\galex{\emph{GALEX}\,}
\def\rosat{\emph{ROSAT}\,}
\def\toi{451}
\def\tid{257605131}
\def\target{TOI 451}

\shorttitle{A Planetary System in Psc--Eri}
\shortauthors{Newton et al. }

\begin{document}

\title{\textit{TESS} Hunt for Young and Maturing Exoplanets (THYME) IV:
Three small planets orbiting a 120 Myr-old star in the Pisces--Eridanus\footnote{sometimes called Meingast-1} stream}
\correspondingauthor{Elisabeth R. Newton}
\email{Elisabeth.R.Newton@Dartmouth.edu}

\author[0000-0003-4150-841X]{Elisabeth R. Newton}
\affiliation{Department of Physics and Astronomy, Dartmouth College, Hanover, NH 03755, USA}

\author[0000-0003-3654-1602]{Andrew W. Mann}
\affiliation{Department of Physics and Astronomy, The University of North Carolina at Chapel Hill, Chapel Hill, NC 27599, USA} 

\author[0000-0001-9811-568X]{Adam L. Kraus}
\affiliation{Department of Astronomy, The University of Texas at Austin, Austin, TX 78712, USA}

\author[0000-0002-4881-3620]{John~H.~Livingston}
\affiliation{Department of Astronomy, University of Tokyo, 7-3-1 Hongo, Bunkyo-ku, Tokyo 113-0033, Japan}

\author[0000-0001-7246-5438]{Andrew Vanderburg}
\altaffiliation{NASA Sagan Fellow}
\affiliation{Department of Astronomy, University of Wisconsin-Madison, Madison, WI 53706, USA}
\affiliation{Department of Astronomy, The University of Texas at Austin, Austin, TX 78712, USA}

\author[0000-0002-2792-134X]{Jason L. Curtis}
\affiliation{Department of Astrophysics, American Museum of Natural History, Central Park West, New York, NY, USA}

\author[0000-0001-5729-6576]{Pa Chia Thao}%
\altaffiliation{NSF GRFP Fellow} 
\affiliation{Department of Physics and Astronomy, The University of North Carolina at Chapel Hill, Chapel Hill, NC 27599, USA} 

\author[0000-0002-1423-2174]{Keith Hawkins}
\affiliation{Department of Astronomy, The University of Texas at Austin, Austin, TX 78712, USA}

\author[0000-0001-7336-7725]{Mackenna L. Wood}%
\affiliation{Department of Physics and Astronomy, The University of North Carolina at Chapel Hill, Chapel Hill, NC 27599, USA} 

\author[0000-0001-9982-1332]{Aaron C. Rizzuto}
\altaffiliation{51 Pegasi b Fellow}
\affiliation{Department of Astronomy, The University of Texas at Austin, Austin, TX 78712, USA}

\author[0000-0002-0345-2147]{Abderahmane Soubkiou}%
\affiliation{Oukaimeden Observatory, PHEA Laboratory, Cadi Ayyad University, BP 2390 Marrakech Morocco}

\author[0000-0003-2053-0749]{Benjamin M. Tofflemire}
\altaffiliation{51 Pegasi b Fellow}
\affiliation{Department of Astronomy, The University of Texas at Austin, Austin, TX 78712, USA}

\author{George Zhou}
\affiliation{Center for Astrophysics $\vert$ Harvard \& Smithsonian, 60 Garden St, Cambridge, MA, 02138, USA}

\author{Ian J.\ M.\ Crossfield}
\affiliation{Department of Physics and Astronomy, University of
  Kansas, Lawrence, KS, USA}

\author[0000-0003-3904-7378]{Logan A.\ Pearce}
\affil{Steward Observatory, University of Arizona, Tucson, AZ 85721, USA}

\author[0000-0001-6588-9574]{Karen A.\ Collins}
\affiliation{Center for Astrophysics $\vert$ Harvard \& Smithsonian, 60 Garden St, Cambridge, MA, 02138, USA}

\author[0000-0003-2239-0567]{Dennis M.\ Conti}
\affiliation{American Association of Variable Star Observers, 49 Bay State Road, Cambridge, MA 02138, USA}

\author[0000-0001-5603-6895]{Thiam-Guan Tan}
\affiliation{Perth Exoplanet Survey Telescope, Perth, Western Australia}

\author{Steven Villeneuva} 
\affiliation{Department of Physics and Kavli Institute for Astrophysics and Space Research, Massachusetts Institute of Technology, Cambridge, MA 02139, USA}

\author[0000-0001-9263-6775]{Alton Spencer}
\affiliation{Western Connecticut State University, Danbury, CT 06810}

\author[0000-0003-2313-467X]{Diana~Dragomir}
\affiliation{Department of Physics and Astronomy, University of New Mexico, 210 Yale Blvd NE, Albuquerque, NM 87106, USA}

\author[0000-0002-8964-8377]{Samuel N. Quinn}
\affiliation{Center for Astrophysics $\vert$ Harvard \& Smithsonian, 60 Garden St, Cambridge, MA, 02138, USA}

\author[0000-0002-4625-7333]{Eric L. N. Jensen}
\affiliation{Dept.\ of Physics \& Astronomy, Swarthmore College, Swarthmore PA 19081, USA}

\author[0000-0003-2781-3207]{Kevin I.\ Collins}
\affiliation{George Mason University, 4400 University Drive, Fairfax, VA, 22030 USA}

\author[0000-0003-2163-1437]{Chris Stockdale}
\affiliation{Hazelwood Observatory, Australia}

\author[0000-0001-5383-9393]{Ryan Cloutier}
\affiliation{Center for Astrophysics $\vert$ Harvard \& Smithsonian, 60 Garden St, Cambridge, MA, 02138, USA}

\author{Coel Hellier}
\affiliation{Astrophysics Group, Keele University, Staffordshire ST5 5BG, U.K.}

\author{Zouhair Benkhaldoun}
\affiliation{Oukaimeden Observatory, PHEA Laboratory, Cadi Ayyad University, BP 2390 Marrakech Morocco}


\author{Carl Ziegler}
\affiliation{Dunlap Institute for Astronomy and Astrophysics, University of Toronto, 50 St. George Street, Toronto, Ontario M5S 3H4, Canada}

\author[0000-0001-7124-4094]{C\'{e}sar Brice\~{n}o}
\affiliation{Cerro Tololo Inter-American Observatory/NSF’s NOIRLab, Casilla 603, La Serena, Chile} 

\author{Nicholas Law}
\affiliation{Department of Physics and Astronomy, The University of North Carolina at Chapel Hill, Chapel Hill, NC 27599, USA}


\author{Bj\"orn Benneke}
\affiliation{Department of Physics and Institute for Research on Exoplanets, Universit\'e de Montréal, Montreal, QC, Canada}

\author[0000-0002-8035-4778]{Jessie L. Christiansen}
\affiliation{Caltech/IPAC-NASA Exoplanet Science Institute, 770 S. Wilson Avenue, Pasadena, CA 91106, USA}

\author[0000-0002-8990-2101]{Varoujan Gorjian}
\affiliation{Caltech/NASA Jet Propulsion Laboratory, 4800 Oak Grove Dr, Pasadena, CA 91109, USA}

\author[0000-0002-7084-0529]{Stephen~R.~Kane}
\affiliation{Department of Earth and Planetary Sciences, University of California, Riverside, CA 92521, USA}

\author{Laura Kreidberg}
\affiliation{Max-Planck-Institut f\"ur Astronomie, K\"onigstuhl 17, 69117 Heidelberg, Germany}

\author[0000-0001-9414-3851]{Farisa Y. Morales}
\affiliation{Jet Propulsion Laboratory,
California Institute of Technology, 4800 Oak Grove Drive, Pasadena, CA 91109, USA}

\author{Michael W Werner} 
\affiliation{Caltech/IPAC-NASA Exoplanet Science Institute, 770 S. Wilson Avenue, Pasadena, CA 91106, USA}

\author[0000-0002-6778-7552]{Joseph D. Twicken}
\affiliation{SETI Institute/NASA Ames Research Center}

\author{Alan~M.~Levine}
\affiliation{Department of Physics and Kavli Institute for Astrophysics and Space Research, Massachusetts Institute of Technology, Cambridge, MA 02139, USA}

\author[0000-0002-5741-3047]{David~ R.~Ciardi}
\affiliation{Caltech/IPAC-NASA Exoplanet Science Institute, 770 S. Wilson Avenue, Pasadena, CA 91106, USA}

\author[0000-0002-5169-9427]{Natalia~M.~Guerrero}
\affiliation{Department of Physics and Kavli Institute for Astrophysics and Space Research, Massachusetts Institute of Technology, Cambridge, MA 02139, USA}

\author{Katharine Hesse}
\affiliation{Department of Physics and Kavli Institute for Astrophysics and Space Research, Massachusetts Institute of Technology, Cambridge, MA 02139, USA}

\author[0000-0003-1309-2904]{Elisa~V.~Quintana}
\affiliation{NASA Goddard Space Flight Center, 8800 Greenbelt Road, Greenbelt, MD 20771, USA}

\author{Bernie Shiao}
\affiliation{Mikulski Archive for Space Telescopes}

\author[0000-0002-6148-7903]{Jeffrey C. Smith}
\affiliation{SETI Institute/NASA Ames Research Center}

\author[0000-0002-5286-0251]{Guillermo~Torres}
\affiliation{Center for Astrophysics $\vert$ Harvard \& Smithsonian, 60 Garden St, Cambridge, MA, 02138, USA}

\author[0000-0003-2058-6662]{George~R.~Ricker}%
\affiliation{Department of Physics and Kavli Institute for Astrophysics and Space Research, Massachusetts Institute of Technology, Cambridge, MA 02139, USA}

\author[0000-0001-6763-6562]{Roland~Vanderspek}%
\affiliation{Department of Physics and Kavli Institute for Astrophysics and Space Research, Massachusetts Institute of Technology, Cambridge, MA 02139, USA}

\author[0000-0002-6892-6948]{Sara~Seager}%
\affiliation{Department of Physics and Kavli Institute for Astrophysics and Space Research, Massachusetts Institute of Technology, Cambridge, MA 02139, USA}
\affiliation{Department of Earth, Atmospheric and Planetary Sciences, Massachusetts Institute of Technology, Cambridge, MA 02139, USA}
\affiliation{Department of Aeronautics and Astronautics, MIT, 77 Massachusetts Avenue, Cambridge, MA 02139, USA}

\author[0000-0002-4265-047X]{Joshua~N.~Winn}%
\affiliation{Department of Astrophysical Sciences, Princeton University, 4 Ivy Lane, Princeton, NJ 08544, USA}

\author[0000-0002-4715-9460]{Jon M. Jenkins}%
\affiliation{NASA Ames Research Center, Moffett Field, CA, 94035, USA}

\author[0000-0001-9911-7388]{David~W.~Latham}
\affiliation{Center for Astrophysics $\vert$ Harvard \& Smithsonian, 60 Garden St, Cambridge, MA, 02138, USA}

\vspace{0.3in}

\begin{abstract}

Young exoplanets can offer insight into the evolution of planetary atmospheres, compositions, and architectures. We present the discovery of the young planetary system TOI \toi\ (TIC \tid, Gaia DR2 4844691297067063424). TOI \toi\ is a member of the 120-Myr-old Pisces--Eridanus stream (Psc--Eri). We confirm membership in the stream with its kinematics, its lithium abundance, and the rotation and UV excesses of both TOI 451 and its wide binary companion, TOI 451 B (itself likely an M dwarf binary). We identified three candidate planets transiting in the \tess\ data and followed up the signals with photometry from \spitzer\ and ground-based telescopes. The system comprises three validated planets at periods of $1.9$, $9.2$ and $16$ days, with radii of $1.9$, $3.1$, and $4.1$ \rearth, respectively. The host star is near-solar mass with $V=11.0$ and $H=9.3$ and displays an infrared excess indicative of a debris disk. The planets offer excellent prospects for transmission spectroscopy with \hst\ and \jwst, providing the opportunity to study planetary atmospheres that may still be in the process of evolving.

\end{abstract}


\section{Introduction}

Exoplanets are expected to undergo significant evolution in the first few hundred million years of their lives, including thermal and compositional changes to their atmospheres and dynamical evolution. Stellar high energy irradiation, which diminishes with age, impacts atmospheric mass loss rates \citep[\eg][]{JacksonCoronal2012, KubyshkinaYoung2018} and atmospheric chemistry \citep[\eg][]{SeguraBiosignatures2005, GaoDeflating2019}. These processes can have a dramatic effect on the observed properties of planets with sizes in between those of Earth and Neptune. 

Atmospheric mass loss is thought to be responsible for the observed ``radius valley'', a deficit of planets $1.5-2$\rearth\ and the accompanying bimodality of the radius distribution \citep{OwenKepler2013, FultonCaliforniaKepler2017}. This valley was predicted by photoevaporation models, where mass loss is driven by high energy radiation from the host star \citep{LopezUnderstanding2013, OwenKepler2013, JinPlanetary2014}. Comparison of models to the data by \citet{OwenEvaporation2017} and \citet{JinCompositional2018} support this interpretation.
However, core-powered mass-loss, in which the atmospheric loss is driven by the luminosity of the hot planetary interior, is also successful at explaining the radius valley \citep{GinzburgCorepowered2018, GuptaSculpting2019}. The timescale for core-powered mass loss is $\sim$1~Gyr \citep{GuptaSignatures2020}, in contrast to $\sim$100~Myr for photoevaporation \citep{OwenEvaporation2017}.
Alternatively, \citet{ZengGrowth2019} and \citet{MousisIrradiated2020} propose that the 2--4~\rearth\ planets are water worlds, with compositions reflecting the planets' accretion and migration history. \cite{2020arXiv200801105L} consider formation in gas-poor environments to argue that the radius valley is primordial.

Planets larger than $\sim1.6$\rearth\ are expected to have gas envelopes constituting $\gtrsim1$\% of the core mass \citep{RogersMost2015, WolfgangHow2015}; and their atmospheric compositions and chemistry can be probed with transmission spectroscopy \citep[\eg][]{SeagerTheoretical2000, Miller-RicciAtmospheric2009}. The observed spectra of these planets range from flat and featureless \citep[][]{KnutsonFeatureless2014, KreidbergClouds2014} to exhibiting the spectral fingerprints of water \citep[][]{FraineWater2014, BennekeWater2019, TsiarasWater2019}. Featureless spectra may result from clouds or hazes present at low atmospheric pressures \citep[\eg][]{MorleyThermal2015}. \cite{GaoDeflating2019} quantified how hazes can also result in large optical depths at low pressures (high altitudes), which results in larger planetary radii than would otherwise be measured. \cite{GaoDeflating2019} find that this effect would be most important in young, warm, and low-mass exoplanets, for which outflows result in high altitude hazes.

The theories make different predictions about atmospheric properties and the timescale for changes. Therefore, the compositions and atmospheric properties of individual young exoplanets, and the distribution of young planet radii can constrain these theories. Transmission spectroscopy of young planets may also allow atmosphere composition measurements where older planets yield flat spectra, if relevant atmospheric dynamics or chemistry change with time.

The \textit{Transiting Exoplanet Survey Satellite} \citep[\tess;][]{RickerTransiting2015} mission provides the means to search for young exoplanets that orbit stars bright enough for atmospheric characterization and mass measurements. The \tess\ Hunt for Young and Maturing Exoplanets (THYME) Survey seeks to identify planets transiting stars in nearby, young, coeval populations. We have validated three systems to date: DS Tuc A b \citep{NewtonTESS2019}, HIP 67522 b \citep{RizzutoTESS2020}, and HD 63433 b and c \citep{MannTESS2020}. Our work complements the efforts of other groups to discover young exoplanets, such as the Cluster Difference Imaging Photometric Survey \citep[CDIPS;][]{2019ApJS..245...13B} and the PSF-based Approach to TESS High quality data Of Stellar clusters project \citep[PATHOS;][]{2019MNRAS.490.3806N}.

The unprecedented astrometric precision from \gaia\ \citep{GaiaCollaborationGaia2016, GaiaCollaborationGaia2018} and \tess's nearly all-sky coverage combined to create an opportunity for the study of young exoplanets that was not previously available. \citet{MeingastExtended2019} conducted a search for dynamically cold associations in phase space using velocities and positions from \gaia. They identified a hitherto unknown stream extending 120$\degree$ across the sky at a distance of only 130 pc, which was called the Pisces--Eridanus stream (Psc--Eri)  by \citet{CurtisTESS2019}. 
\citet{MeingastExtended2019} found that the 256 sources defined a main sequence, and based on the presence of a triple system composed of three giant stars, they suggested an age of $\sim$1~Gyr.
\citet{CurtisTESS2019} extracted \tess\ light curves for a subset of members and measured their rotation periods. Finding that the stellar temperature--period distribution closely matches that of the Pleiades at 120 Myr, they determined that the stream is similarly young. Color--magnitude diagrams from \citet{CurtisTESS2019}, 
\citet{RoserCensus2020}, and \citet{RatzenbockExtended2020}, and 
lithium abundances from \citet{Arancibia-SilvaLithiumrotation2020} and \citet{HawkinsChemical2020}, support the young age. 

The Psc--Eri stream offers a new set of young, nearby stars around which to search for planets. The stream complements the similarly-aged Pleiades, in which no exoplanets have been found to date. Thanks to the nearly all-sky coverage of \tess, photometry that could support a search for planets orbiting Psc–Eri members was already available when the stream was identified. We cross-matched the \tess\ Objects of Interest \citep{GuerreroTESS2020} alerts\footnote{Now the TOI releases; \url{https://tess.mit.edu/toi-releases/}} to the list of Psc--Eri members from \citet{CurtisTESS2019}, and found TOI 451 to be a candidate member of the stream.\footnote{As noted in \citet{CurtisTESS2019}.} 

We present validation of three planets around TOI 451 with periods of $1.9$ d (TOI 451 b), $9.2$ d (TOI 451 c) and $16$ d (TOI 451 d). In Section \ref{sec:data}, we present our photometric and spectroscopic observations. In Section \ref{sec:measure}, we discuss our measurements of the basic parameters and rotational properties of the star. We find that TOI 451 is a young solar-mass star, and has a comoving companion, Gaia DR2 4844691297067064576, that we call TOI 451 B. We address membership of TOI 451 and TOI 451 B to the Psc--Eri stream in Section \ref{sec:member} through kinematics, abundances, stellar rotation, and activity. We model the planetary transits seen by \tess, \spitzer, PEST, and LCO in Section \ref{sec:analysis}, and conclude in Section \ref{sec:summary}. Appendix \ref{appendix} describes our analysis of \galex\ and demonstrates UV excess as a way to identify new low-mass members of Psc--Eri.

\section{Observations}\label{sec:data}

\subsection{Time-series photometry}

\begin{figure*}
    \centering
    \includegraphics[width=0.9\textwidth]{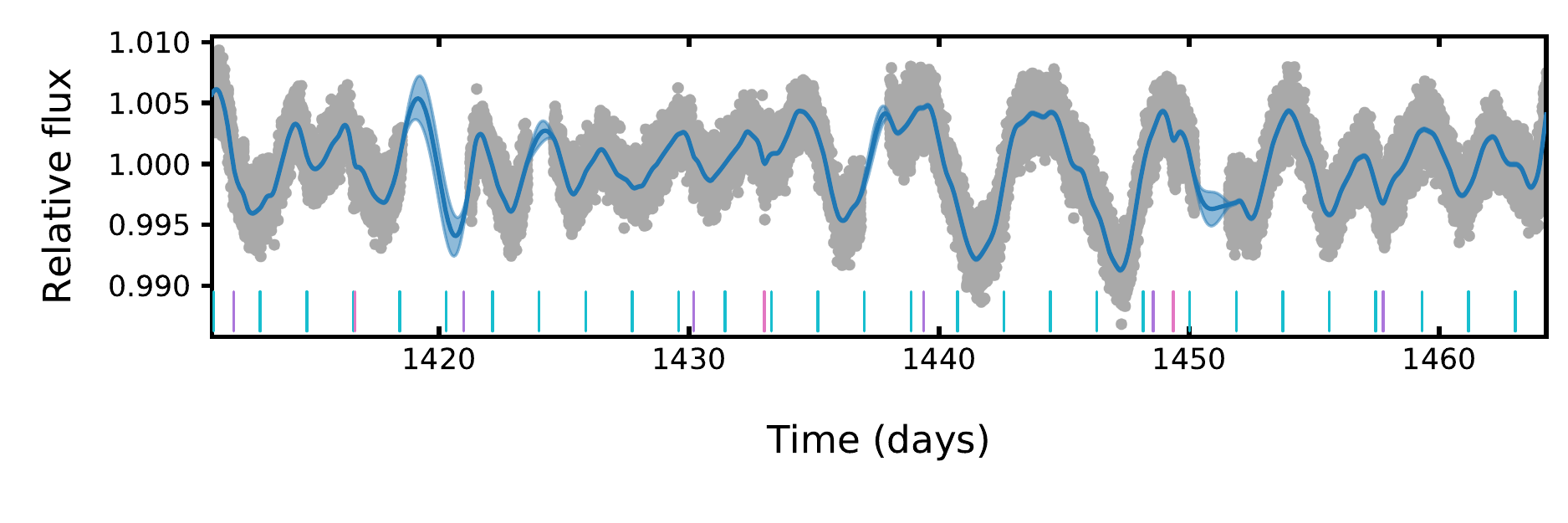}
    \includegraphics[width=0.9\textwidth]{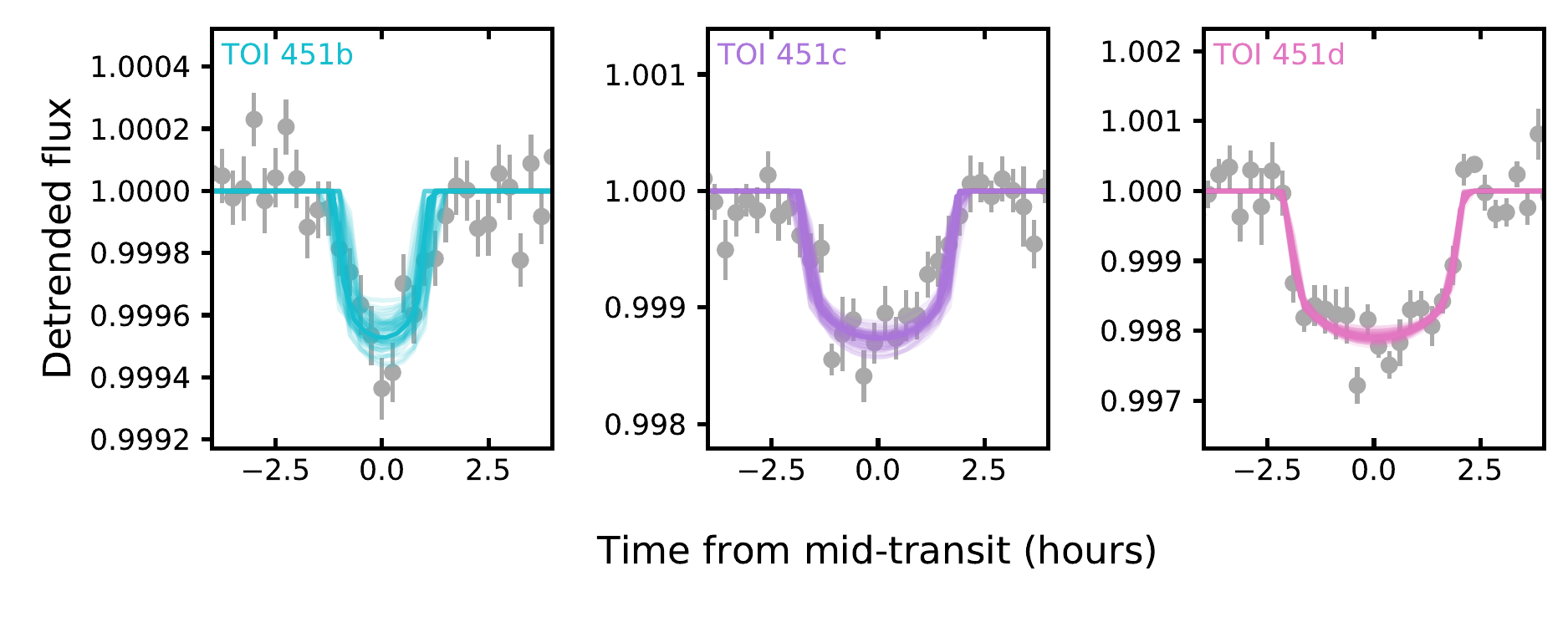}
    \caption{\tess\ data and models. {\bf Top panel}: \tess\ lightcurve (gray points) with the Gaussian process (GP) model to stellar variability overlain (blue). The mean (opaque line) and 68\% confidence limits (semi-transparent regions) of the stellar variability model are shown. Marked at the bottom are transit times of each of the three planets, with TOI 451 b in teal, c in purple, and d in pink. {\bf Bottom panel}: the phase-folded data centered on the transit of each planet, after the best-fit stellar variability model has been removed and the transits of other planets have been masked. The best-fit transit models (opaque lines) are overplotted along with 50 draws from the posterior distribution (semi-transparent lines). Both the data and the models have been binned into 15 minute bins. Note that small changes in the binning will cause the perceived shape of the plotted transit data to change.}
    \label{fig:tess}
\end{figure*}

\begin{figure*}
    \centering
    \includegraphics[width=0.9\textwidth]{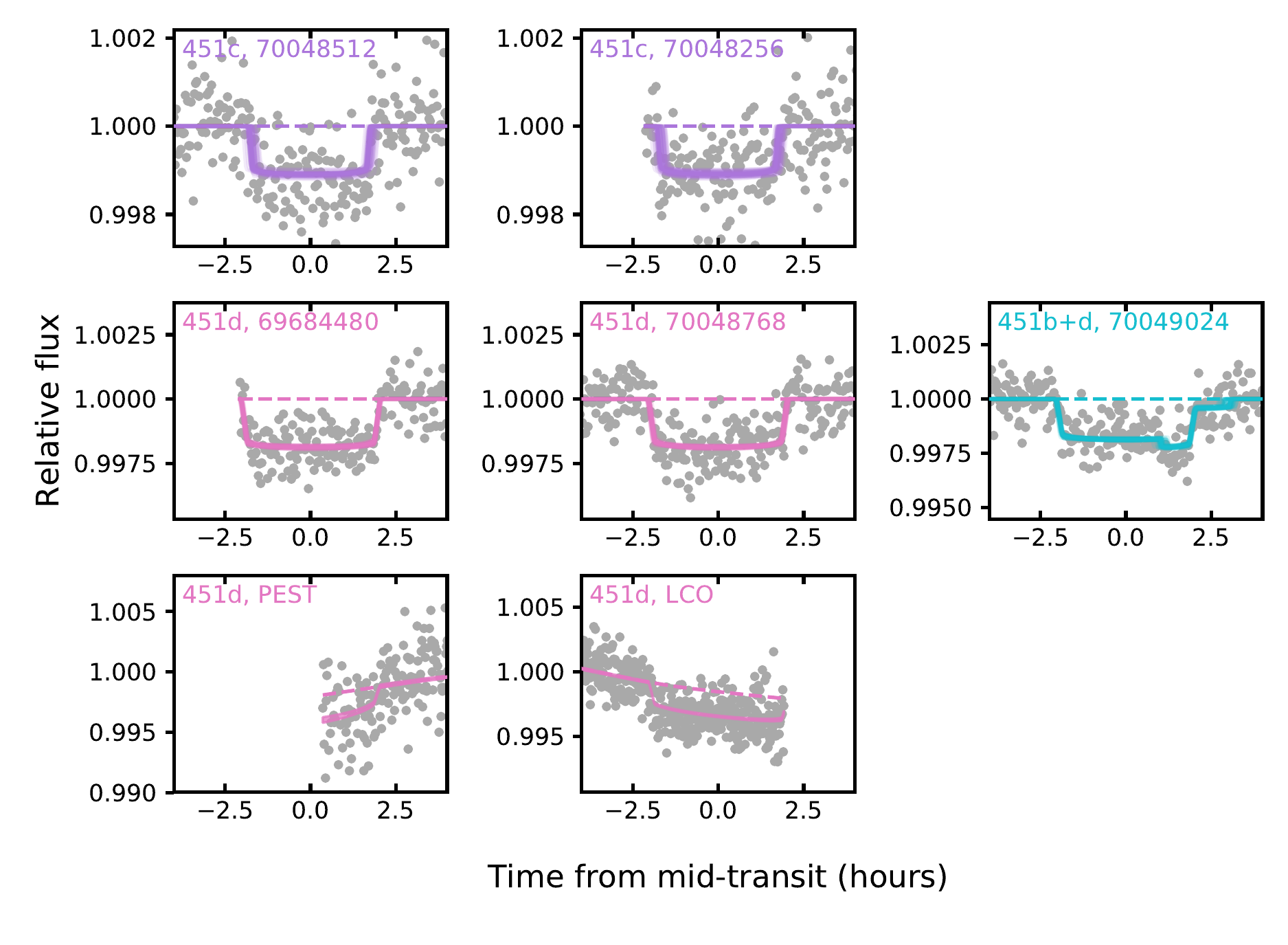}
    \caption{Follow-up photometry (gray points) and model (solid lines). The model without the transit is also shown (dashed lines). The model shown is the best-fit transit model for the case where eccentricities are fixed to 0. 50 draws from the posterior are also overplotted. {\bf Top two rows:} \spitzer\ data and models.  No stellar variability is modeled for \spitzer\ transits. {\bf Bottom row:} Ground-based observatory data and models, with PEST on the left and LCO in the center. Stellar variability is modeled for these data.}
    \label{fig:other}
\end{figure*}

After the initial discovery of the signal in  the \tess\ data (Figure \ref{fig:tess}), we obtained follow-up transit photometry from the Las Cumbres Observatory \citep[LCO;][]{BrownCumbres2013}, the Perth Exoplanet Survey Telescope (PEST), and the \spitzer\ Space Telescope\footnote{May it orbit in peace.} (Figure \ref{fig:other}). Table \ref{tab:photometry} lists the photometric data modeled in Section \ref{sec:modeling}; this subset of the available data provided the best opportunity for constraining the transit model. These and additional ground-based lightcurves\footnote{https://exofop.ipac.caltech.edu/tess/target.php?id=257605131} ruled out eclipsing binaries on nearby stars, found the transit to be achromatic, and confirmed TOI 451 as the source of the three candidate event.

\begin{deluxetable}{lllll} 
\tablecaption{Photometry used in this paper \label{tab:photometry} } 
\tablehead{ \colhead{Telescope} & \colhead{Filter} & \colhead{Date} & \colhead{Planet} & \colhead{AOR}}
\startdata
    \tess       &   \tess   &     2018-10-19     &  \ldots &  \ldots \\
                &           &     --2018-12-11    &  \ldots &  \ldots \\
    PEST        &   $Rc$    &     2019-11-05    & d &  \ldots \\
    LCO-CTIO    &   $z_s$   &     2019-12-08    & d &  \ldots \\
    \spitzer    &   Ch2     &     2019-06-11    & d & 69684480 \\
                &           &     2019-12-24    & d+b & 70049024 \\
                &           &     2019-12-26    & c & 70048512 \\
                &           &     2020-01-09    & d & 70048768 \\
                &           &     2020-01-13    & c & 70048256
\smallskip 
\enddata 
\tablecomments{In the text, \spitzer\ transits are referred to be the AOR ID (last column). \spitzer\ AOR 69684480 obtained through GO program 14084 (PI: Crossfield); remaining AORs through GO program 14011 (PI: Newton). }
\end{deluxetable}

\subsubsection{\textit{TESS}}

\tess\ observed TIC \tid\ in Sectors 4 and 5, from 2018 October 19 to 2018 December 11. TIC \tid\ was identified as a promising target to support \tess's prime mission by \citet{StassunTESS2018}; it was included in the Candidate Target List (CTL) and thus observed at 2 minute cadence. The data were processed by the Science Processing and Operations Center (SPOC) pipeline \citep{JenkinsTESS2016}, which calibrated and extracted the data, corrected the lightcurve, and finally filtered the light curve and searched for planets. The identified signal passed visual vetting, and the community was notified via alerts along with other \tess\ candidate planets \citep{GuerreroTESS2020}. 

The original \tess\ alert identified one candidate, TOI 451.01, at a period of $8$ days. Our initial exploration with \texttt{EXOFAST} \citep{EastmanEXOFAST2013} suggested that the $8$ day signal is a combination of two real but distinct transit signals, and yielded two candidates, at $9.2$ and $16$ days. Throughout follow-up, these signals were referred to as TOI 451.02 and 451.01, respectively. Another candidate was added as a community TOI to ExoFOP, identified as TIC 257605131.02, at a period of $1.9$ days. The candidate was also identified in the results of two independent pipelines run by our team. We now identify the $1.9$-day planet as TOI 451\,b, the $9.2$-day as TOI 451\,c and the $16$-day as TOI 451\,d.

The presence of all three planets is supported by the \tess\ data. Our custom pipeline \citep{RizzutoZodiacal2017, RizzutoTESS2020} identified all three planets, with signal-to-noise ratio (S/N) $= 14.8$, $13.9$, and $10.4$ (for planet detection, we typically adopt a threshold of S/N $=7$). TOI 451\,d was also found at the correct period in the SPOC multi-sector transit search. The transit detection statistic was $12.2$. It had a clean data validation report \citep{2018PASP..130f4502T,2019PASP..131b4506L} and passed the odd-even depth test, the difference image centroiding and ghost diagnostic test, which can reveal background eclipsing binaries. The only diagnostic test it failed was the statistical bootstrap test, which was due to the transits of the other planets in the light curve that were not identified.  TOI 451 c narrowly missed detection with a transit detection statistic of $7.05$.

We used the presearch data conditioning simple aperture photometry (\texttt{PDCSAP\_FLUX}) light curve produced by the SPOC pipeline \citep{StumpeKepler2012,SmithKepler2012, StumpeMultiscale2014}. Prior to using these data for our transit fit, we removed flares by iteratively fitting the Gaussian Process model described in \S\ref{sec:analysis} using least-squares regression. We first masked the transits and iterated the GP fitting, rejecting outliers at each of three iterations. We then detrended the lightcurve using the fitted GP model, and removed outliers from the detrended lightcurve. We removed $3.5\sigma$ outliers and iterated until no additional points were removed, removing a total of 47 data points.

\subsubsection{\textit{Spitzer}}\label{sec:spitzer}

We obtained transit observations of TOI 451 c and d at $4.5\micron$ (channel 2) with \spitzer's Infrared Array Camera \citep[IRAC;][]{FazioInfrared2004}, one on 2019 June 11 (UT) and four between 2019 December 24 and 2020 January 13 (UT). Dates and AOR designations, which we use to identify the \spitzer\ transits, are in Table \ref{tab:photometry}.

We used $2$ second frames and the $32\times32$ pixel subarray. We placed the target on the detector ``sweet spot'' and used the ``peak-up'' pointing mode to ensure precise pointing \citep{IngallsIntrapixel2012, IngallsRepeatability2016}. For AORs 70049024, 70048512, 70048768, and 70048256, we scheduled a 20 min dither, then an 8.5 hr stare covering the transit, followed by a 10 minute dither.  For AOR 69684480, we scheduled a 30 min dither, 8.7 hr stare, and 15 min dither. Though we had not considered  TOI 451 b at the time of scheduling \spitzer\ observations, a transit of TOI 451 b happened to coincide with one of our transits of TOI 451 d.

We extracted time-series photometry and pixel data from the {\it Spitzer} AORs\footnote{Available from \url{https://sha.ipac.caltech.edu/applications/Spitzer/SHA/}} following the procedure described in \citet{Livingston:etal:2019b}. Apertures of 2.2-2.4 pixels were used, selected based on the algorithm to minimize both red and white noise also described in that work. We used pixel-level decorrelation \citep[PLD;][]{DemingSpitzer2015} to model the systematics in the {\it Spitzer} light curves, which are caused by intra-pixel sensitivity variations coupled with pointing jitter. We used the \texttt{exoplanet} package \citep{exoplanet:exoplanet} to jointly model the transit and systematics in each {\it Spitzer} light curve, assuming Gaussian flux errors.\footnote{\texttt{exoplanet} uses \texttt{starry} \citep{exoplanet:luger18} to efficiently compute transit models with quadratic limb darkening under the transformation of \citet{KippingEfficient2013}, and estimates model parameters and uncertainties via \texttt{theano} \citep{exoplanet:theano} and \texttt{PyMC3} \citep{exoplanet:pymc3}.} 
We assume a circular orbit. For the prior on the limb-darkening parameters, we used the values tabulated by \citet{ClaretGravity2011} in accordance to the stellar parameters. We placed priors on the stellar density, planetary orbital period, planet-to-star radius ratio ($R_{p}/R_{s}$), and expected time of transit based on an initial fit to only the {\it TESS} data. For all planetary parameters, the priors were wider than the posterior distributions, so the data provides the primary constraint.
We obtained initial maximum a posteriori (MAP) parameter estimates via the gradient-based {\tt BFGS} algorithm \citep{NoceWrig06} implemented in {\tt scipy.optimize}. We then explored parameter space around the MAP solution via the {\tt NUTS} Hamiltonian Monte Carlo sampler \citep{Hoffman:Gelman:2014} implemented in {\tt PyMC3}. We confirmed the posteriors were unimodal and updated the MAP estimate if a higher probability solution was found.

The two transits of c (AORs 70048512 and 70048256) and solely of d (AORs 69684480 and 70048768) were fit simultaneously.
For AOR 70049024, the model consisted of overlapping transits of TOI 451 b and d, and we fixed the limb-darkening parameters. Trends were included where visual inspection of the model components showed that PLD alone was not sufficient to explain the data, i.e. due to stellar variability; for AORs 69684480 and 70048512, quadratic terms were used.
We then computed PLD-corrected light curves by subtracting the systematics model corresponding to the MAP sample from the data. We use these corrected {\it Spitzer} datasets for the subsequent transit analyses in \S\ref{sec:analysis}.

The inclusion of TOI 451 b overlapping the transit of d in AOR 70049024 was strongly favored by the data per the Bayesian Information Criterion (BIC). The two-planet model had $\Delta$BIC$=27$ compared to the model with only TOI 451 d. Prior to realizing the transit of TOI 451 b was present, we had also considered a model with TOI 451 d and a Gaussian Process. The two-planet model had $\Delta$BIC$=7.7$ compared to this model.

\subsubsection{Investigation into \textit{Spitzer} systematics}

With the joint PLD fit of the two \spitzer\ transits of TOI 451 c, we consistently modeled both ($\rprstar=0.034 \pm 0.001$). However, there was a $5\sigma$ discrepancy between the depths of the two transits when we used PLD as described above, but reduced each independently ($\rprstar = 0.037\pm0.001$ for AOR 70048512; $\rprstar = 0.026\pm0.002$ for AOR 70048256). The difference in the two transit depths, obtained 18.5 days apart, is difficult to explain astrophysically, especially given the expectation for decreased stellar variability at $4.5\micron$. We hypothesize that this results from the low signal-to-noise ratio of the transits, but the discrepancy in the independent fits raised concerns about systematic effects in the \spitzer\ data. 

To investigate systematics in these data further, we additionally used the BiLinearly-Interpolated Subpixel Sensitivity (BLISS) mapping technique to produce an independent reduction of the {\it Spitzer} data for TOI 451\,c. BLISS uses a non-parametric approach to correct for \spitzer's intrapixel sensitivity variations.

We processed the \spitzer\ provided Basic Calibrated Data (BCD) frames using the Photometry for Orbits, Eccentricities, and Transits (POET; \citealt{StevensonTransit2012, 2011ApJ...727..125C}) pipeline to create systematics-corrected light curves. This included masking and flagging bad pixels, and calculating the Barycentric Julian Dates for each frame. The center position of the star was fitted using a two-dimensional, elliptical Gaussian in a 15 pixel square window centered on the target’s peak pixel. Simple aperture photometry was performed using a radius of 2.5 pixels, an inner sky annulus of 7 pixels, and an outer sky annulus of 15 pixels. 

To correct for the position-dependent (intra-pixel) and time-dependent (ramp) \spitzer\ systematics, we used the BLISS Mapping Technique, provided through POET. We used the most recent 4.5$\mu$m intra-pixel sensitivity map from \cite{MayIntroducing2020}. The transit was modeled using the \cite{MandelAnalytic2002} transit model and three different ramp parameterizations: linear, quadratic, and rising exponential, as well as a no-ramp model. The time-dependent component of the model consisted of the mid-transit time, $R_{p}/R_{s}$, orbital inclination ($\cos{i}$), semi-major axis ratio, system flux, ramp phase, ramp amplitude, and ramp constant offset. These parameters were explored with an Markov Chain Monte Carlo (MCMC) process, using 4 walkers with 500,000 steps and a burn-in region of 1,000 steps. The period was fixed to 9.19 days based on the {\it TESS} data, and starting locations for the MCMC fit were based on test runs.

To determine the best ramp models, we used two metrics: 1) overall minimal red noise levels in the fit residual, assessed by considering the root-mean-squared (rms) binned residuals as a function of different bin sizes with the theoretical uncorrelated white noise; and 2) Bayesian Information Criterion \citep[BIC; e.g.,][]{2014ApJ...797...42C}. Low rms and low BIC are favored.

For AOR 70048512, significant red noise remained for the no-ramp model and was present to a lesser degree with the other ramp models. However, the no-ramp model yielded the lowest BIC value, while quadratic and rising exponential ramps yielded the largest but had lower red noise. There was a discrepancy in the transit depth for the different ramp parameterizations: no-ramp and the linear ramp returned $\rprstar$ of $0.031\pm0.001$ and $0.032\pm0.001$, respectively, while the quadratic and exponential ramps each returned $0.036\pm0.001$ (in comparison to the PLD independent fit of this transit with $\rprstar = 0.037\pm0.001$). 

For AOR 70048256, we obtained consistent transit depths of around $0.032$ with all four BLISS ramp options with typical error $\pm0.002$ (in comparison to the PLD independent fit of this transit with $\rprstar = 0.027\pm0.003$). No significant red noise was present in any of the ramp models. The no-ramp model yielded the lowest BIC value, while the quadratic and rising exponential ramp models yielded the largest. 

This exploration demonstrated that the transit depths were sensitive to ramp choice and the metric used to select the best fit (e.g., lowest BIC, lowest red noise, best agreement with {\it TESS}). We  conclude that additional systematic errors in the {\it Spitzer} transit depths are likely present and are not accounted for in the analysis presented in \S\ref{sec:modeling}.

\subsubsection{LCO}

We observed a transit of TOI 451d in Pan-STARRS $z$-short band with the LCO \citep{BrownCumbres2013} 1\,m network node at Cerro Tololo Inter-American Observatory on the night of 2019 December 08. 
We used the {\tt TESS Transit Finder}, which is a customised version of the {\tt Tapir} software package \citep{Jensen:2013}, to schedule our transit observation. 
The images were calibrated by the standard LCOGT BANZAI pipeline \citep{McCully:2018} and the photometric data were extracted using the {\tt AstroImageJ} ({\tt AIJ}) software package \citep{Collins:2017}. We used a circular aperture with radius 12 pixels to extract differential photometry. The images have stellar point-spread-functions (PSFs) with FWHM $\sim 2\farcs5$.

\subsubsection{PEST}
We observed a transit egress of TOI 451d in the $Rc$-band with PEST on 2019 Nov 05 (UT). PEST is a 12 inch Meade LX200 SCT Schmidt-Cassegrain telescope equipped with a SBIG ST-8XME camera located in a suburb of Perth, Australia. We used a custom pipeline based on C-Munipack\footnote{http://c-munipack.sourceforge.net} to calibrate the images and extract the differential time-series photometry. The transiting event was detected using a $7\farcs4$ aperture centered on the target star. The images have typical PSFs with a FWHM of $\sim4\arcsec$.

\subsubsection{WASP-South}
 WASP-South is an array of 8 cameras located in Sutherland, South Africa. It is the Southern station of the WASP transit-search project \citep{2006PASP..118.1407P}. WASP-South observed available fields with a typical 10-min cadence on each clear night. Until 2012, it used 200-mm, f/1.8 lenses with a broad $V$+$R$ filter, and then switched to 85-mm, f/1.2 lenses with an SDSS-$r$ filter.  TOI 451 was observed for 150 nights in each of 2006, 2007 and 2011 (12\,600 data points) and for 170 nights in each of 2012, 2013 and 2014 (51\,000 data points).  

\subsection{Spectroscopy}

We obtained spectra with the Southern Astrophysical Research (SOAR) telescope/Goodman, South African Extremely Large Telescope (SALT)/High Resolution Spectrograph (HRS), Las Cumbres Observatory (LCO)/Network of Robotic Echelle Spectrographs (NRES), and Small and Moderate Aperture Research Telescope System (SMARTS)/CHIRON. The Goodman spectrum was used to fit the spectral energy distribution (\S\ref{sec:SED}); and the SALT, LCO, and CHIRON spectra to measure radial velocities (RVs; \S\ref{Sec:rvs}). The S/N of the spectra used for RVs are given in Table \ref{tab:rvs}.

\subsubsection{SOAR/Goodman}

We obtained a spectrum of \target\ with the Goodman High-Throughput Spectrograph \citep{Goodman} on the Southern Astrophysical Research (SOAR) 4.1 m telescope located at Cerro Pachón, Chile. On 2019 Dec 3 (UT), we took five exposures of \target 
with the red camera, the 1200 l/mm grating in the M5 setup, and the 0.46\arcsec\ slit rotated to the parallactic angle. This setup yielded a resolution of $R \simeq 5900$ spanning 6250--7500\AA. To account for drifts in the wavelength solution, we obtained Ne arc lamp exposures throughout the night. We took standard calibration data (dome/quartz flats and biases) during the preceeding afternoon.

We performed bias subtraction, flat fielding, optimal extraction of the target spectrum, and found the wavelength solution using a 4th-order polynomial derived from the Ne lamp data. We then stacked the five extracted spectra using the robust weighted mean (for outlier removal). The stacked spectrum had S/N$>100$ over the full observed wavelength range. 

\subsubsection{SALT/HRS}

We obtained six epochs with HRS \citep{CrausePerformance2014} on SALT \citep{BuckleyCompletion2006} between 2019 July and 2019 October. Each epoch consisted of three back-to-back exposures. We used the high-resolution mode, ultimately obtaining an effective resolution of 46,000. Flat-fielding and wavelength calibration were performed using the MIDAS pipeline \citep{KniazevMN482016,KniazevSALT2017}.

\subsection{LCO/NRES}

We observed \target\ twice using NRES \citep[][]{SiverdNRES18} on the LCO system. NRES is a set of cross-dispersed echelle spectrographs connected to 1\,m telescopes within the Las Cumbres network, providing a resolving power of $R=53,000$ over the range $3800-8600$ \AA. We took both observations at the Cerro Tololo node, the first on 2019 March 27 and the second on 2019 Jul 31 (UT). The March observation consisted of two back-to-back exposures. 
The standard NRES pipeline\footnote{\url{https://lco.global/documentation/data/nres-pipeline/}} reduced extracted, and wavelength calibrated both observations.

\subsection{SMARTS/CHIRON}

We obtained a single spectrum with the CHIRON spectrograph \citep{TokovininCHIRON2013} on SMARTS, from which we measured the radial velocity and rotational broadneing of the star. CHIRON is a $R=80,000$ high resolution fiber bundle fed spectrograph on the 1.5\,m SMARTS telescope, located at located at Cerro Tololo Inter-American Observatory (CTIO), Chile. Data were reduced according to \citet{TokovininCHIRON2013}.

\subsection{Speckle Imaging}

\begin{figure}
    \centering
    \includegraphics[width=0.49\textwidth]{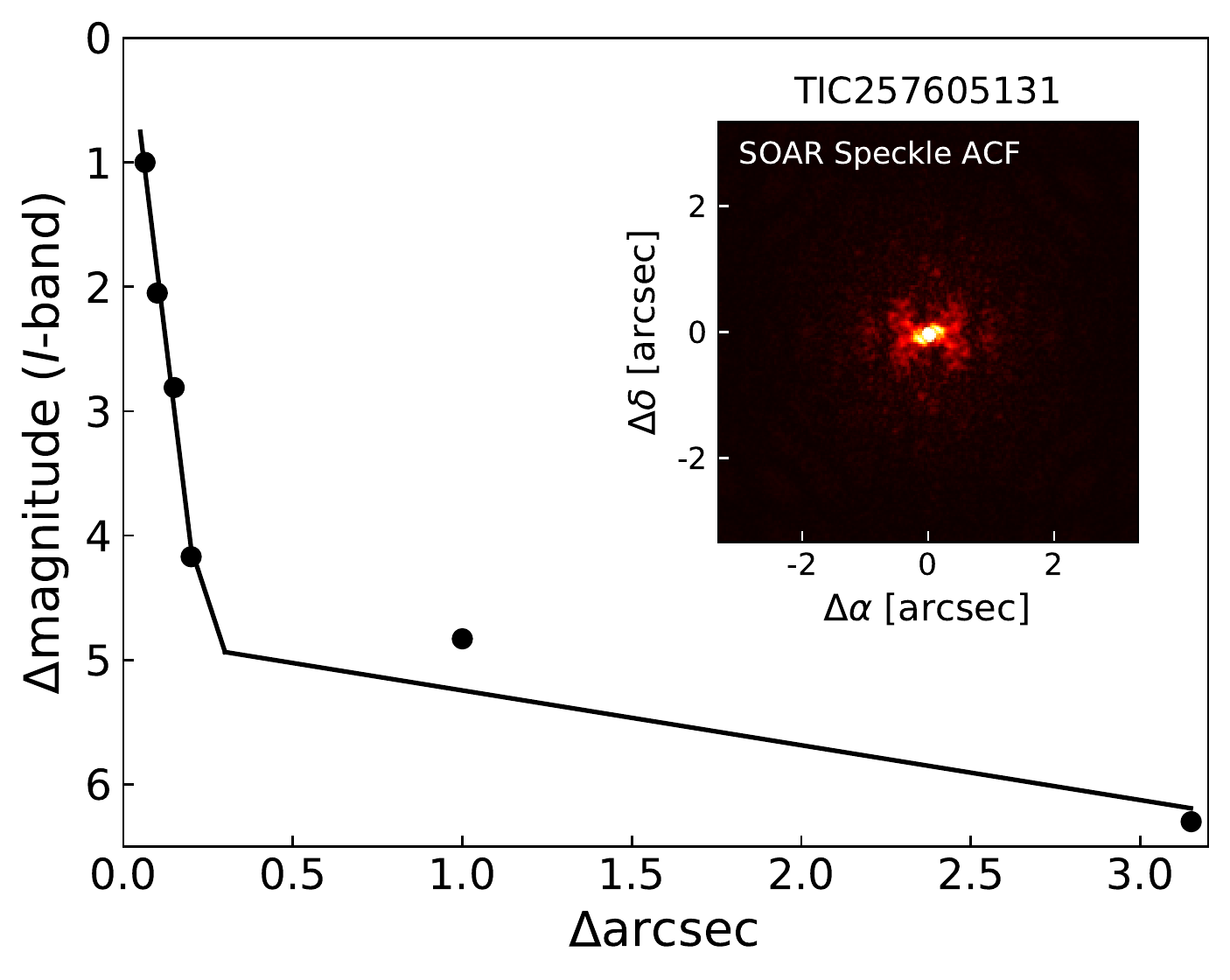}
    \caption{Detection limits (5$\sigma$) for companions to TOI 451 from SOAR speckle imaging.}
    \label{fig:soar}
\end{figure}

To rule out unresolved companions that might impact our interpretation of the transit, we obtained speckle imaging using the High-Resolution  Camera (HRCam) on the SOAR telescope. We searched for sources near \target\ in SOAR speckle imaging obtained on 17 March 2019 UT in $I$-band, a similar visible bandpass as \tess. Further details of observations from the SOAR \tess\ survey are available in \cite{ZieglerSOAR2020}.  We detected no nearby stars within $3\arcsec$ of \target\ within the $5\sigma$ detection sensitivity of the observation (Figure ~\ref{fig:soar}).

\subsection{Gaia astrometry}

\subsubsection{A wide binary companion}

The Gaia DR2 catalog \citep{GaiaCollaborationGaia2018} includes one co-moving, co-distant neighbor at $\rho = 37.8$\arcsec\, ($\rho = 4700$ AU), TIC 257605132 (Gaia DR2 4844691297067064576). Aside from entries in all-sky catalogs, this neighbor star is unremarkable and does not appear to have been previously studied in the astronomical literature. Though this companion is only about two {\it TESS} pixels away from TOI 451, our high spatial resolution {\it Spitzer} data definitively rule it out as the source of the transits.

The parallax difference between TOI 451 and its neighbor is consistent with zero at 1.2$\sigma$, and the relative velocity in the plane of the sky ($\Delta \mu = 0.21 \pm 0.09$ mas/yr; $\Delta v_\mathrm{tan} = 0.12 \pm 0.05$ km/s) is lower than the circular orbital velocity for a total pair mass of $M \sim 1.2 M_{\odot}$ and a projected distance of $4700$ AU. The relative astrometry and kinematics are thus consistent with a bound binary system and much lower than the typical velocity dispersion of $\sim$1 km/s seen in young associations. The separation is also within the semi-major axis range range commonly seen for bound binary pairs among young low-density associations \citep{2008ApJ...686L.111K} and the field solar-type binary distribution \citep{RaghavanSurvey2010}. We therefore concluded that TIC 257605132 is a bound binary companion to TOI 451, and hereafter refer to it as TOI 451 B.

TOI 451 B has {\it Gaia} DR2 parameters of $B_P-R_P=2.527$ mag and $T_\mathrm{eff}=3507$ K. The color corresponds to a spectral type of M3V according to \citet{KimanExploring2019}. TOI 451 B itself is likely a binary, as its Renormalized Unit Weight Error \citep[RUWE;][]{GaiaDr2}\footnote{\url{https://gea.esac.esa.int/archive/documentation/GDR2/Gaia_archive/chap_datamodel/sec_dm_main_tables/ssec_dm_ruwe.html}} is $RUWE = 1.24$, higher than the distribution typically seen for single stars and indicative of binarity (\citealt{zeit8}; Kraus et al., in preparation). We compared TOI 451 B's location in the $G-R_P$ vs. $M_G$ color--magnitude diagram to the similarly-aged Pleiades population from \citet{Lodieu5D2019}; it lies 0.65 mag above the main-sequence locus. We assumed TOI 451 B is comprised of two near-equal mass stars, and adjusted the reported 2MASS $K$ magnitude ($K=10.76$ mag) by 0.65 mag to match the main sequence locus. We then applied the mass-$M_K$ relation from \citet{MannHow2019}, which resulted in a mass of $0.45$ \msun. This mass is in agreement with the expectations for a star of the observed \textit{Gaia} DR2 color for TOI 451 B.

\subsubsection{Limits on Additional Companions}\label{sec:comp}

Our null detection from speckle interferometry is consistent with the deeper limits set by the lack of {\it Gaia} excess noise. TOI 451 has $RUWE = 0.91$, consistent with the distribution of values seen for single stars. Based on a calibration of the companion parameter space that would induce excess noise (\citealt{zeit8}; \citealt{BelokurovUnresolved2020}; Kraus et al., in preparation), this corresponds to contrast limits of $\Delta G \sim 0$ mag at $\rho = 30$ mas, $\Delta G \sim 4$ mag at $\rho = 80$ mas, and $\Delta G \sim 5$ mag at $\rho \ge 200$ mas. Given an age of $\tau = 120$ Myr at $D = 124$ pc, the evolutionary models of \citet{bhac15} would imply corresponding physical limits for equal-mass companions at $\rho \sim 4$ AU, $M \sim 0.45 M_{\odot}$ at $\rho \sim 10$ AU, and $M \sim 0.30 M_{\odot}$ at $\rho > 25$ AU.

Other than TOI 451 B, Gaia DR2 does not report any other comoving, codistant neighbors within an angular separation of $\rho < 600\arcmin$ ($\rho < 75,000$ AU) from TOI 451. At separations beyond this limit, any neighbor would be more likely to be an unbound member within a loose unbound association like Psc-Eri, rather than a bound binary companion \citep{2008ApJ...686L.111K}, so we concluded that there are no other wide bound binary companions to TOI 451 above the Gaia catalog's completeness limit. The lack of any such companions at $40 < \rho < 600$\arcsec\, also further supports that TOI 451 B is indeed a bound companion and not a chance alignment with another unbound Psc-Eri member, as the local sky density of Psc-Eri members appears to be quite low ($\Sigma \la 3 \times 10^{-3}$ stars arcmin$^{-2}$) given the lack of other nearby stars in \gaia. 

\cite{ZieglerMeasuring2018} and \cite{BrandekerContrast2019} have mapped the completeness limit close to bright stars to be $\Delta G \sim 6$ mag at $\rho = 2\arcsec$, $\Delta G \sim 8$ mag at $\rho = 3\arcsec$, and $\Delta G \sim 10$ mag at $\rho = 6\arcsec$. The evolutionary models of \citet{bhac15} would imply corresponding physical limits of $M \sim 0.20 M_{\odot}$ at $\rho = 250$ AU, $M \sim 0.085 M_{\odot}$ at $\rho = 375$ AU, and $M \sim 0.050 M_{\odot}$ at $\rho = 750$ AU. At wider separations, the completeness limit of the Gaia catalog \citep[$G \sim 20.5$ mag at moderate galactic latitudes;][]{GaiaCollaborationGaia2018} corresponds to an absence of any companions down to a limit of $M \sim 0.050 M_{\odot}$.

\subsection{Literature photometry}

We gathered optical and near-infrared (NIR) photometry from the literature for use in our determination of the stellar parameters. Optical photometry comes from \gaia\ DR2 \citep[][]{Evans2018,GaiaDr2}, AAVSO All-Sky Photometric Survey \citep[APASS,][]{2012JAVSO..40..430H}, and SkyMapper \citep{Skymapper1}. NIR photometry comes from the Two-Micron All-Sky Survey \citep[2MASS,][]{Skrutskie2006}, and the Wide-field Infrared Survey Explorer \citep[WISE,][]{Wright2010}. 

\section{Measurements}\label{sec:measure}

\begin{deluxetable}{lccc}
\centering
\tabletypesize{\scriptsize}
\tablewidth{0pt}
\tablecaption{Properties of the host star \target. \label{tab:params}}
\tablehead{\colhead{Parameter} & \colhead{Value} & \colhead{Source} }
\startdata
\multicolumn{3}{c}{Identifiers}\\
\hline
TOI & 451 & \\ 
TIC & 257605131 &   \\
TYC & 7577-172-1 &  \\
2MASS & J04115194-3756232 & \\
Gaia DR2 & 4844691297067063424 &  \\
\hline
\multicolumn{3}{c}{Astrometry}\\
\hline
$\alpha$  & 04 11 51.947 & \emph{Gaia} DR2\\
$\delta$  & -37 56 23.22 & \emph{Gaia} DR2 \\
$\mu_\alpha$ (mas\,yr$^{-1}$)&-11.167$\pm$0.039	& \emph{Gaia} DR2\\
$\mu_\delta$  (mas\,yr$^{-1}$) & 12.374$\pm$0.054  & \emph{Gaia} DR2\\
$\pi$ (mas) & 8.0527$\pm$0.0250& \emph{Gaia} DR2\\
\hline
\multicolumn{3}{c}{Photometry}\\
\hline
G$_{Gaia}$ (mag) & $10.7498\pm0.0008$& \emph{Gaia} DR2\\
BP$_{Gaia}$ (mag) & $11.1474\pm0.0027$  & \emph{Gaia} DR2\\
RP$_{Gaia}$ (mag) & $10.2199\pm0.0017$ & \emph{Gaia} DR2\\
B$_T$ (mag) & $11.797\pm0.074$ & Tycho-2 \\%
V$_T$ (mag) &  $11.018\pm0.064$  & Tycho-2\\%
J (mag) & $ 9.636\pm0.024 $ &  2MASS\\
H (mag) & $9.287\pm0.022 $ & 2MASS\\	
$K_S$ (mag) & $9.190\pm0.023 $ & 2MASS\\
W1 (mag) &$9.137\pm0.024$  & ALLWISE\\%
W2 (mag)& $9.173\pm0.020$ & ALLWISE\\
W3 (mag)& $9.117\pm0.027 $ & ALLWISE\\ 
W4 (mag)& $ 8.632\pm0.292$  & ALLWISE\\ 
\hline
\multicolumn{3}{c}{Kinematics \& Position}\\
\hline
Barycentric RV (km\, s$^{-1}$) & $ 19.87\pm0.12  $ & This paper \\
Distance (pc) & 123.74$\pm$0.39 & \citet{gaia_distances}\\
U (km\, s$^{-1}$) & $-10.92\pm0.05 $ & This paper\\ 
V (km\, s$^{-1}$) & $-4.18\pm0.08$ & This paper\\
W (km\, s$^{-1}$) & $-18.81\pm0.09$ & This paper\\ 
X (pc) & $-41.56\pm0.14$ & This paper\\%
Y (pc) & $-73.61\pm0.24$ & This paper\\
Z (pc) & $ -90.43\pm0.29$ & This paper\\
\hline
\multicolumn{3}{c}{Physical Properties}\\
\hline
Rotation Period (days) & $5.1\pm0.1$ d  & This paper\\
\vsini (km\, s$^{-1}$) & $7.9\pm0.5$ \kms & This paper\\
$i_*$ ($^\circ$) & $69^{11}_{-8} {\degree}$ & This paper\\
\fbol\,(erg\,cm$^{-2}$\,s$^{-1}$)& ($1.23\pm0.07)\times10^{-8}$ & This paper\\ 
T$_{\mathrm{eff}}$ (K) & $5550 \pm 56$ & This paper\\
M$_\star$ (M$_\odot$) & $ 0.950\pm0.020 $ & This paper \\ 
R$_\star$ (R$_\odot$) &  $0.879 \pm 0.032$ & This paper \\
L$_\star$ (L$_\odot$) & $0.647 \pm 0.032$ & This paper \\
$\rho_\star$ ($\rho_\odot$) & $1.4 \pm 0.16$ & This paper \\
Age (Myr) & $125\pm8$ & \cite{StaufferKeck1998}$^a$ \\
& $112\pm5$ & \cite{2015ApJ...813..108D}$^a$ \\
& $120$ & \cite{CurtisTESS2019} \\
& $134\pm6.5$ & \cite{RoserCensus2020} \\
E(B-V) (mag) & $0.02^{+0.04}_{-0.01}$ & This paper \\
\enddata
\tablecomments{Age references denoted $^a$ are ages for the Pleiades. $i_*$ adopts the convention $i_*<90 \degree$}
\end{deluxetable}

\subsection{Stellar parameters}\label{Sec:stellarparam}

We summarize our derived stellar parameters in Table~\ref{tab:params}. 

\subsubsection{Luminosity, effective temperature, and radius}\label{sec:SED}
To determine $L_*$, \teff\, and $R_*$ of \target, we simultaneously fit its spectral energy distribution using the photometry listed in Table~\ref{tab:params}, our SOAR/Goodman optical spectrum, and Phoenix BT-Settl models \citep{Allard2011}. Significantly more detail of the method can be found in \citet{Mann2015b} for nearby un-reddened stars, with details on including interstellar extinction in \citep{Mann2016b}. 

We compared the photometry to synthetic magnitudes computed from our SOAR spectrum. We used a Phoenix BT-Settl model \citep{Allard2011} to cover gaps in the spectra and simultaneously fitting for the best-fitting Phoenix model and a reddening term (since reddening impacts both the spectrum and photometry). The Goodman spectrum is not as precisely flux-calibrated as the data used in \citet{Mann2015b}, so we included two additional free parameters to fit out wavelength-dependent flux variations. The bolometric flux of \target\ is the integral of the unreddened spectrum. This flux and the Gaia DR2 distance yield an estimate of $L_*$.

\begin{figure}
    \centering
    \includegraphics[width=0.49\textwidth]{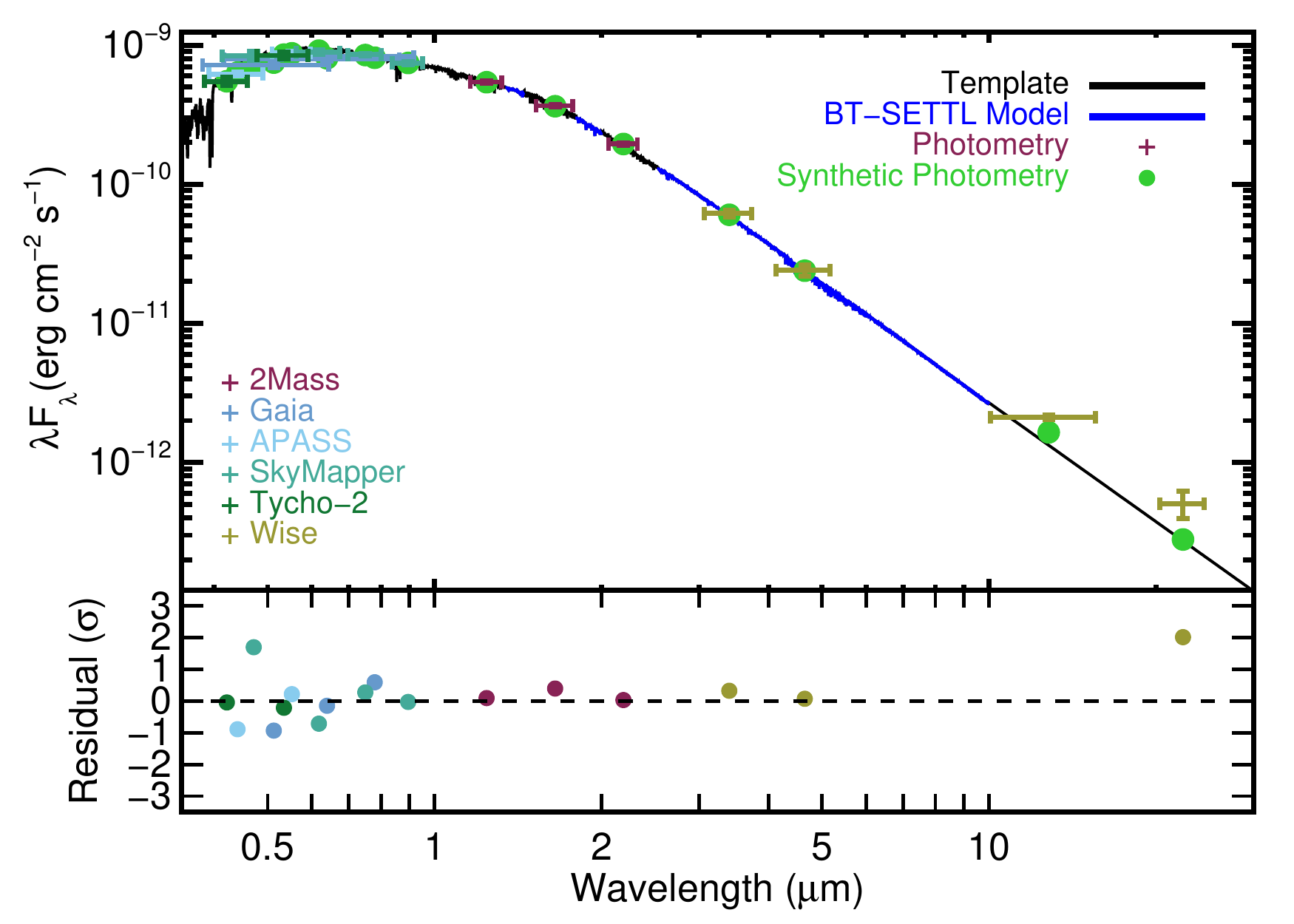}
    \caption{Best-fit spectral template and Goodman spectrum (black) compared to the photometry of TOI 451. Blue regions are BT-SETTL models, used to fill in gaps or regions of high telluric contamination. Literature photometry is colored according to the source with horizontal errors corresponding to the filter width and vertical errors the measurement errors. Corresponding synthetic photometry is shown as green points. The bottom panel shows the residuals in terms of standard deviations from the fit, with a single point ($W1$) off the scale.}
    \label{fig:sed}
\end{figure}

We show the best-fit result in Figure~\ref{fig:sed} and adopted stellar parameters in Table~\ref{tab:params}. Our fitting resulted in two consistent radius estimates: the first from the Stefan-Boltzmann relation (with \teff\ from the model grid) and the second from the $R_*^2/$(distance)$^2$ scaling (i.e., how much the BT-Settl model needs to be scaled to match the absolutely-calibrated spectrum). The latter method is similar to the infrared-flux method \citep[IRFM,][]{Blackwell1977}. Both measurements depend on a common parallax and observed spectrum, and hence are not completely independent. However, the good agreement ($<$1$\sigma$) was a useful confirmation of the final fit. Our derived parameters were \teff$=5550\pm56$\,K, $L_*=0.647\pm0.032L_\odot$, $R_*=0.879\pm0.032R_\odot$ from Stefan-Boltzmann and $R_*=0.863\pm0.024R_\odot$ from the IRFM. We adopt the former $R_*$ for all analyses for consistency with previous work. 

\subsection{Infrared excess}

The two reddest bands, $W3$ (12\um) and $W4$ (22\um) were both brighter than the derived from the best-fit template. We estimated an excess flux of $26\pm7$\% at $W3$, and $70\pm30\%$ at $W4$ ($3.7$ and $2.3\sigma$, respectively, assuming Gaussian errors). This suggests a cool $\lesssim300\,K$ debris disk. The $W2$ excess was not significant, both because of the large uncertainty in the $W4$ magnitude (8.63$\pm$0.29) and because the SED analysis is sensitive to the choice of template at long wavelengths. 

The frequency of infrared excesses decreases with age, declining from tens of percent at ages less than a few hundred Myr to a few percent in the field \citep{2008ApJ...673L.181M, 2007ApJ...654..580S, 2009ApJS..181..197C}. In the similarly-aged Pleiades cluster, \spitzer\ 24\um\ excesses are seen in 10\% of FGK stars \citep{2006ApJ...649.1028G}. This excess emission suggests the presence of a debris disk, in which planetesimals are continuously ground into dust \citep[see][for a review]{2018ARA&A..56..541H}.

\subsubsection{Mass}
To determine the mass of \target, we used the MESA Isochrones and Stellar Tracks \citep[MIST; ][]{DotterMESA2016, ChoiMESA2016a}. We compared all available photometry to the model-predicted values, accounting for errors in the photometric zero-points, reddening, and stellar variability. We restricted the comparison to stellar ages of 50--200\,Myr and solar metallicity based on the properties of the stream. We assumed Gaussian errors on the magnitudes, but included a free parameter to describe underestimated uncertainties in the models or data. The best-fit parameters from the MIST models were $M_*=0.950\pm0.020 M_\odot$, $R_*=0.850\pm0.015R_\odot$, \teff$=5555\pm$45\,K, and $L_*=0.610\pm0.030L_\odot$. These were consistent with our other determinations, but we adopt our empirical $L_*$, \teff\ and $R_*$ estimates from the SED and only utilize the $M_*$ value from the evolutionary models in our analysis.

\subsection{Radial velocities}\label{Sec:rvs}

We used high resolution optical spectra from SALT/HRS, NRES/LCO, and SMARTS/CHIRON to determine stellar radial velocities (RVs). We did not include Gaia because the RV zero-point has not been established in the same manner as our ground-based data. 

We computed the spectral-line broadening functions \citep[BFs;][]{Rucinski1992,Tofflemireetal2019} through linear inversion of our spectra with a narrow-lined template. For the template, we used a synthetic PHOENIX model with \teff\ $=5400$ K and \logg\ $= 4.5$ \citep{2013A&A...553A...6H}. The BF accounts for the RV shift and line broadening. We computed the BF for each echelle order and combined them weighted by their S/N. We then fit a Gaussian profile to the combined BF to measure the RV. The RV uncertainty was determined from the standard deviation of the best-fit RV from three independent subsets of the echelle orders. 

For HRS epochs, which consisted of three individual exposures, and the first NRES epoch, which consisted of two individual exposures, the RV and its uncertainty were determined from the error-weighted mean and standard error of the three individual spectra. 

The resulting RVs are listed in Table \ref{tab:rvs}. The RV zero-points were calculated from the spectra obtained in this work and \cite{RizzutoTESS2020} and are based on telluric features. The zero-points are: $0.05\pm0.10$ for HRS (28 spectra), $ 0.32 \pm 0.09$ for NRES (11 spectra), and $-0.05\pm0.16$ for CHIRON (1 spectrum). The S/N was assessed at $\sim 6580\AA$. 

\begin{deluxetable}{l r c c c}
\tablecaption{Radial velocity measurements of TOI 451 \label{tab:rvs}}
\tablewidth{0pt}
\tablehead{
\colhead{Site} & \colhead{BJD} & \colhead{RV} & \colhead{$\sigma_{RV}$} & \colhead{S/N}\\
\colhead{} & \colhead{} & \colhead{(km s$^{-1}$)} & \colhead{(km s$^{-1}$)} & \colhead{}
}
\startdata
HRS	 	 &	2458690.652	& 19.8	& 0.1 & 80\\
HRS	 	 &	2458705.606	& 19.7	& 0.1 & 86\\
HRS	 	 &	2458709.606	& 19.8	& 0.1 & 70\\
HRS	 	 &	2458713.584	& 19.9	& 0.1 & 60\\
HRS	 	 &	2458752.482	& 20.04	& 0.03 & 83\\
HRS	 	 &	2458760.468	& 20.0	& 0.2 & 74\\
NRES         &	2458695.868	& 20.20	& 0.09 & 13\\
NRES         &	2458699.850	& 20.3	& 0.1 & 7\\
CHIRON	 	 &	2458529.560	& 20.00	& 0.06 & 24\\
\hline
\multicolumn{4}{l}{Weighted mean: 19.9 (km/s)} \\
\multicolumn{4}{l}{RMS: 0.12 (km/s)} \\
\multicolumn{4}{l}{Std Error: 0.04 (km/s)} \\
\enddata
\tablecomments{The zero-points are {\it not} included in the individual velocities but are accounted for in the statistics listed in the bottom rows. The zero-points are: $0.05\pm0.10$ for HRS, $ 0.32 \pm 0.09$ for NRES, and $-0.05\pm0.16$ for CHIRON. The S/N was assessed at $\sim 6580\AA$. For the HRS and NRES epochs, which consist of multiple back-to-back spectra, the S/N for the middle spectrum is listed.}
\end{deluxetable}
%

\subsection{Orbit of TOI 451 and TOI 451 B}

We used Linear Orbits for the Impatient via the python package \texttt{lofti\_gaiaDR2} \citep[LOFTI;][]{PearceOrbital2020} to constrain the orbit of TOI 451 and TOI 451 B. Briefly, the \texttt{lofti\_gaiaDR2} retrieves observational constraints for the components from the \textit{Gaia} archive and fits Keplerian orbital elements to the relative motion using the Orbits for the Impatient rejection sampling algorithm \citep{2017AJ....153..229B}.  Unresolved binaries such as TOI 451 B can pose issues for LOFTI. While TOI 451 B's $RUWE$ suggests binarity, it is still relatively low ($1.2$) and only on the edge of where astrometric accuracy compromises LOFTI \citep{PearceOrbital2020}. We applied LOFTI to the system but caution that the unresolved binary may influence the results to an unknown, but likely small, degree. Our LOFTI fit constrained the orbit of TOI 451 and TOI 451 B to be close to edge-on: we found an orbital inclination of $i=93.8 \pm 11.6 ^\circ$.

\subsection{Stellar rotation}\label{sec:rotation}

\subsubsection{Projected rotation velocity}

We used the high resolution CHIRON spectrum to measure the projected rotation velocity of TOI 451. We deconvolved the observed spectrum against a non-rotating synthetic spectral template from the ATLAS9 atmosphere models \citep{Castelli2004} via a least-squares deconvolution \citep[following][]{Donati1997}. We fitted the line profile with a convolution of components accounting for the rotational, macroturbulent, and instrumental broadening terms. The rotational kernel and the radial tangential macroturbulent kernels were computed as prescribed in \citet{Gray2005}, while the instrument broadening term was a Gaussian of FWHM $3.75$ \kms\ set by the CHIRON resolution. We found a projected rotational broadening of $\vsini = 7.9 \pm 0.5\,\kms$ and a macroturbulence of $2.2 \pm 0.5\,\kms$ for TOI 451. 

\subsubsection{Rotation period}

\begin{figure}
    \centering
    \includegraphics[width=0.49\textwidth]{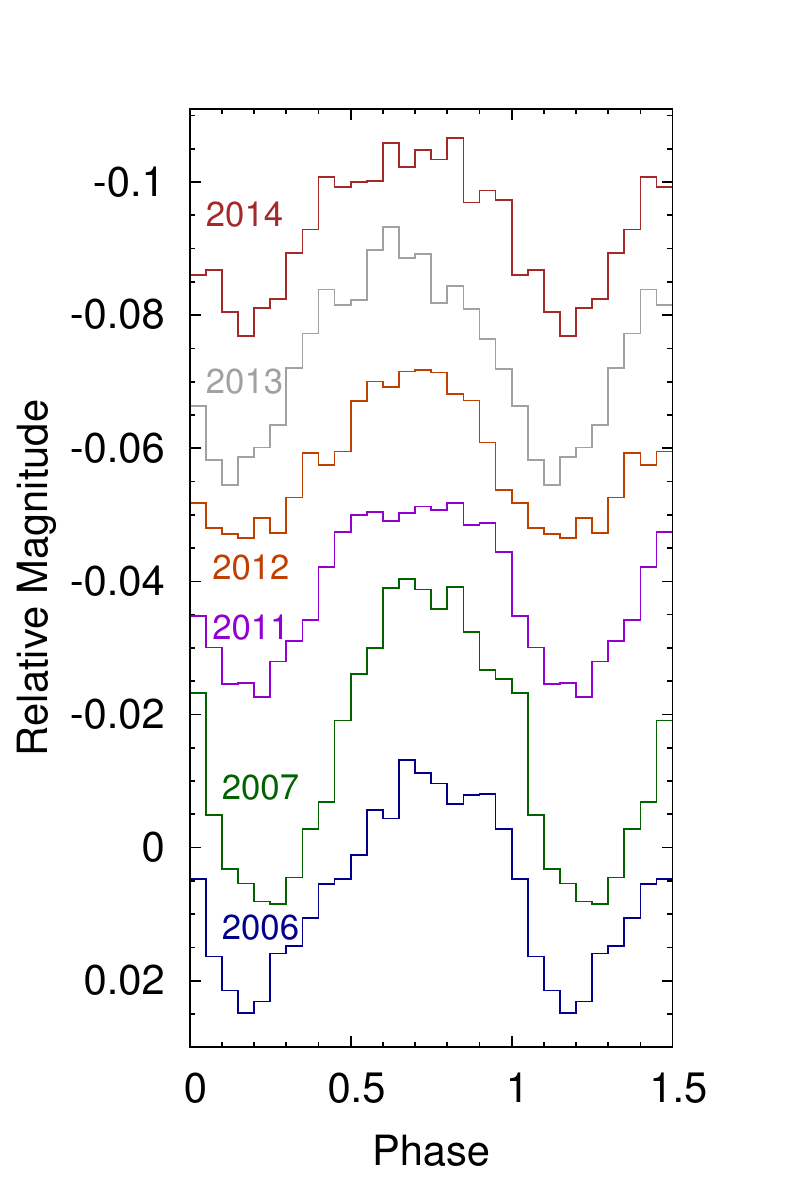}
    \caption{The WASP-South photometry from each year of observation, folded on the 5.2 d rotation period. For display purposes we have added magnitude and phase offsets.}
    \label{fig:wasp}
\end{figure}

In WASP-South, each of the six seasons of data shows clear modulation at 5.2 days, with variations in phase and the amplitude varying from 0.01 to 0.023 (Figure~\ref{fig:wasp}). The mean period from WASP-South is 5.20 days and the standard deviation is 0.02.

We measured the stellar rotation period from the \tess\ data using Gaussian processes (GPs) \citep{AngusInferring2018} as implemented in {\tt celerite} \citep{Foreman-MackeyFast2017}. We used the same rotation kernel as in \citet{NewtonTESS2019}, which is composed of a mixture of two stochastically driven, damped harmonic oscillators. The primary signal is an oscillator at the stellar rotation period $P_*$. The secondary signal is at half the rotation period. We also included a jitter term. The parameters we fit for are described in detail in \S\ref{sec:modeling}, where the GP was fit simultaneously with the transits. We used the period from WASP-South to place a wide prior on the GP fit (see \S\ref{sec:modeling}). The period we measured from the GP is $\ln{P_*}=1.635^{+0.027}_{-0.024}$ or $5.1\pm0.1$ d. 

\subsubsection{Stellar inclination}

We used the procedure outlined in \cite{MasudaInference2020} to infer the inclination $i$ from $P_*$, $\vsini$, $R_*$, and their respective errors. 
This implementation is accurate even in the case of large uncertainties on the rotation period and $\vsini$. We used an affine invariant Markov Chain Monte Carlo sampler \citep{GoodmanENSEMBLE2010} to determine the posterior probability distribution of $\cos{i_*}$. 
We explored the parameter space of $\cos{i_*}$, $R_*$, and $P_{*}$, comparing the $\vsini$ derived from the fit parameters at each step to the measured $\vsini$. We imposed Gaussian priors on $R_*$ and $P_{*}$ based on our measurements for the system, and a prior that $0\leq \cos{i_*}\leq1$.
We took the median and 68\% confidence intervals as the best value and error. Adopting the convention that $i_*<90 {\degree}$, $i_*=69^{+11}_{-8} {\degree}$. 
The $2\sigma$ confidence interval spans 56$\degree$ to 86$\degree$, so the stellar inclination is not inconsistent with alignment between the stellar spin axis, the planetary orbital axes, and the binary orbital axis.

\section{Membership in Psc--Eri}\label{sec:member}

In this section, we present evidence to support identification of TOI 451 as a member of Psc--Eri. While kinematics provide strong support, the stream membership is under active discussion in the literature, so here we consider other indicators of youth.

\subsection{Kinematics}

The original sample from \citet{MeingastExtended2019} required RVs from {\it Gaia} for membership. \citet{CurtisTESS2019} extended the sample to two dozen hotter stars by incorporating RVs from the literature. Recent searches have identified candidate Psc--Eri members without RVs. \cite{RoserCensus2020} adapted the convergent point method \citep{LeeuwenParallaxes2009} to the highly elongated structure of the stream. After placing distance and tangential velocity constraints on stars in the vicinity of the \citet{MeingastExtended2019} sample, they identified 1387 probable stream members. \citet{RatzenbockExtended2020} identified around 2000 new members with a machine learning classifier, trained on the originally identified sample of stream members. \citet{RoserCensus2020} calculated a bulk Galactic velocity for the Psc--Eri members identified by \citet{MeingastExtended2019}  of $(U,V,W)=(-8.84, -4.06, -18.33)\pm(2.2,1.3,1.7)$\,km/s. This agrees with the value we have calculated for \target{} of $(-10.92,-4.18,-18.81)$\,km/s within 1$\sigma$. Based on the space velocities and using the Bayesian membership selection of \citet{2011MNRAS.416.3108R}, we computed a Psc--Eri membership probability of 97\% for \target, and 84\% for the companion \target\ B.

TOI 451 was included (as Gaia DR2 4844691297067063424) in the original membership list from \citet[][]{MeingastExtended2019} and in the subset with roxtation periods from \tess\ data identified by \citet{CurtisTESS2019}. It also was listed as a member in \citet{RatzenbockExtended2020} and \cite{RoserCensus2020}.

\subsection{Atmospheric Parameters and Abundances}
We used the high-resolution (R$\sim$ 46,000) spectra obtained with the SALT telescope to derive stellar     abundances, as well as \teff\ and \logg. The spectra cover $\sim$3700--8900~\AA. We median stacked the spectra to obtain a final spectrum with a S/N around 170 in the continuum at $\sim$5000~\AA. We derived the \teff, \logg, [Fe/H] and the microturbulent velocity (v$_{\mathrm{micro}}$) using the  Brussels Automatic Code for Characterizing High accUracy Spectra \citep[\texttt{BACCHUS};][]{Masseron2016} following the method detailed in \cite{HawkinsChemical2020}. To summarize, we setup BACCHUS using the atomic line list from the fifth version of the Gaia-ESO linelist (Heiter et al., submitted) and molecular information for the following species were also included: CH \citep{Masseron2014}; CN, NH, OH, MgH and  C$_{2}$ (T. Masseron, private communication); and SiH (Kurucz linelists\footnote{http://kurucz.harvard.edu/linelists/linesmol/}). We employed the MARCS model atmosphere grid \citep{Gustafsson2008}, and the TURBOSPECTRUM \citep{Plez2012} radiative transfer code. \texttt{BACCHUS} uses the standard Fe excitation-ionization balance technique to derive \teff, \logg, and [Fe/H]. We refer the reader to Section~3 of \cite{Hawkins2020a} for a more detailed description of \texttt{BACCHUS}. The stellar atmospheric parameters derived from Fe excitation-ionization balance are \teff\ $= 5556 \pm 60$, \logg\ $= 4.62 \pm 0.17$~dex, [Fe/H] $= -0.02 \pm 0.08$~dex. The \teff\ and implied \logg\ derived from a simultaneous fit of the star's spectral-energy-distribution are \teff\ $= 5550 \pm 56$, \logg\ $= 4.53 \pm 0.04$. Encouragingly, these values are, within the uncertainties, consistent with the physical properties outlined in Table~\ref{tab:params}, which were determined without high-resolution spectra. 

Once the stellar atmospheric parameters were determined, we determined the abundance of Li at 6708~\AA, A(Li). The presence (or absence) of large amounts of Li is an age indicator. Li fuses at the relatively low temperature of 2.5 $\times 10^6$~K. As a star ages, Li mixes downward into regions hotter than this temperature, where it is burned into heavier elements. Therefore, the abundance of Li decreases as the star ages. The amount of depletion varies with mass (or \teff, given a  main sequence population). 
Therefore, A(Li) at a given \teff\ constrains a star's age. This applies to both the Galactic disk \citep[e.g.][]{Ramirez2012} and open clusters \citep[e.g.][]{Boesgaard1998, Takeda2013, Martin2018, Bouvier2018}. 

To measure A(Li), we used the \texttt{BACCHUS} module \texttt{abund}. Using \texttt{abund}, we generated a set of synthetic spectra at 6708~\AA\ with differing atmospheric abundances. We then used $\chi^2$ minimization to find the synthetic spectrum that best fits the observed spectrum. We determined A(Li) $= 2.80 \pm 0.10$~dex.

\begin{figure}
    \centering
    \includegraphics[width=0.49\textwidth]{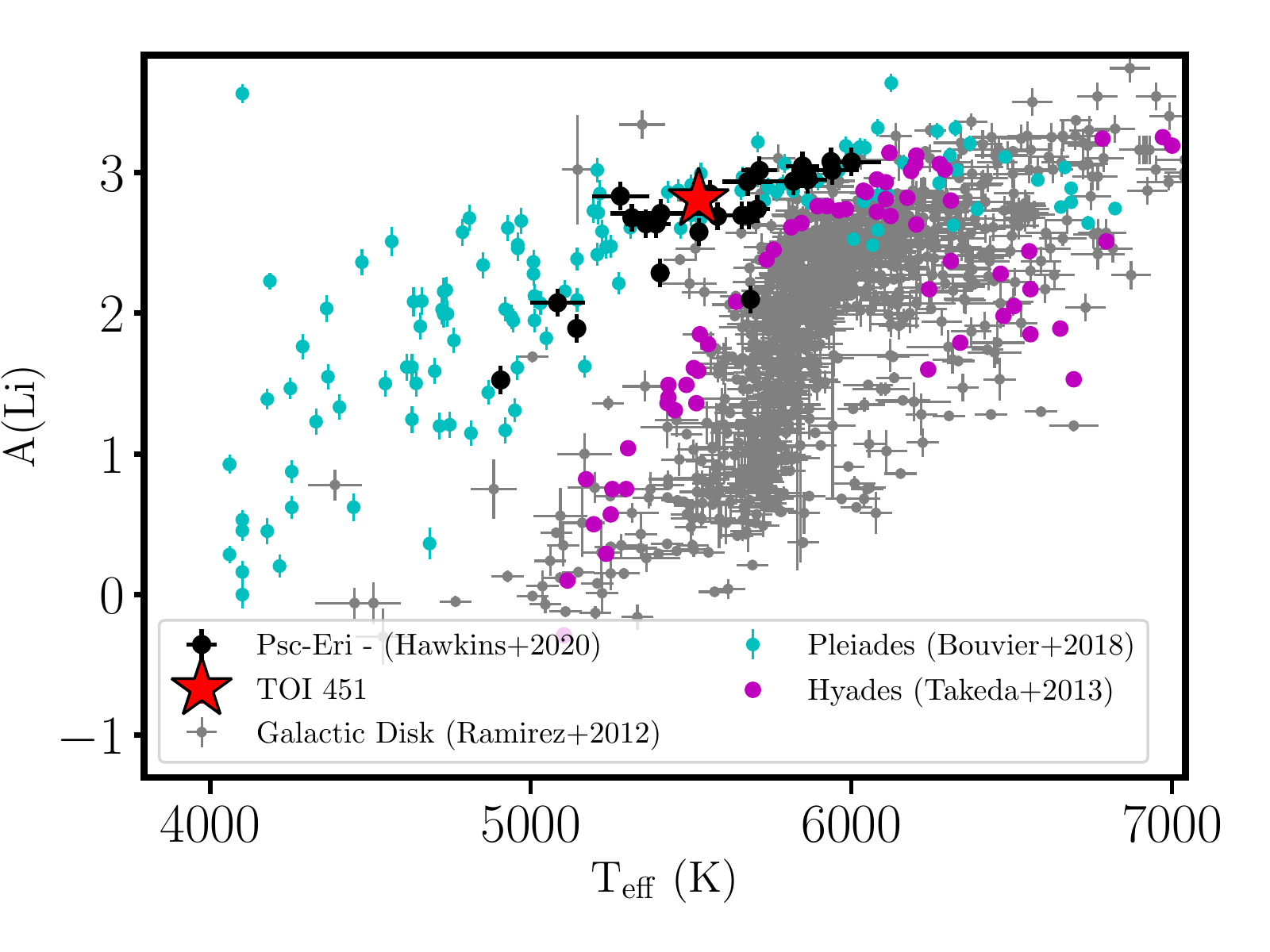}
    \caption{\teff\ as a function of the atmospheric abundance of Li, A(Li), for TOI~451 (red star), the Galactic Disk \citep[gray circles;][]{Ramirez2012}, the Pleiades \citep[cyan circles;][]{Bouvier2018}, the Hyades \citep[magenta circles;][]{Takeda2013}, and the Psc--Eri stream \citep[black circles;][]{HawkinsChemical2020}. TOI 451 has a measured Li abundance that is consistent with the Psc--Eri stream. }
    \label{fig:li}
\end{figure}

We compare A(Li) and \teff\ of TOI 451 to the observed trends  in the Galactic Disk \citep{Ramirez2012}, the Pleiades \citep{Bouvier2018}, and the Hyades \citep{Takeda2013} (Fig.~\ref{fig:li}). A(Li) for TOI~451 closely matches A(Li) for the Psc-Eri stream, indicating that it is likely $\sim$120~Myr old and a stream member.

\begin{figure}
    \centering
    \includegraphics[width=0.49\textwidth]{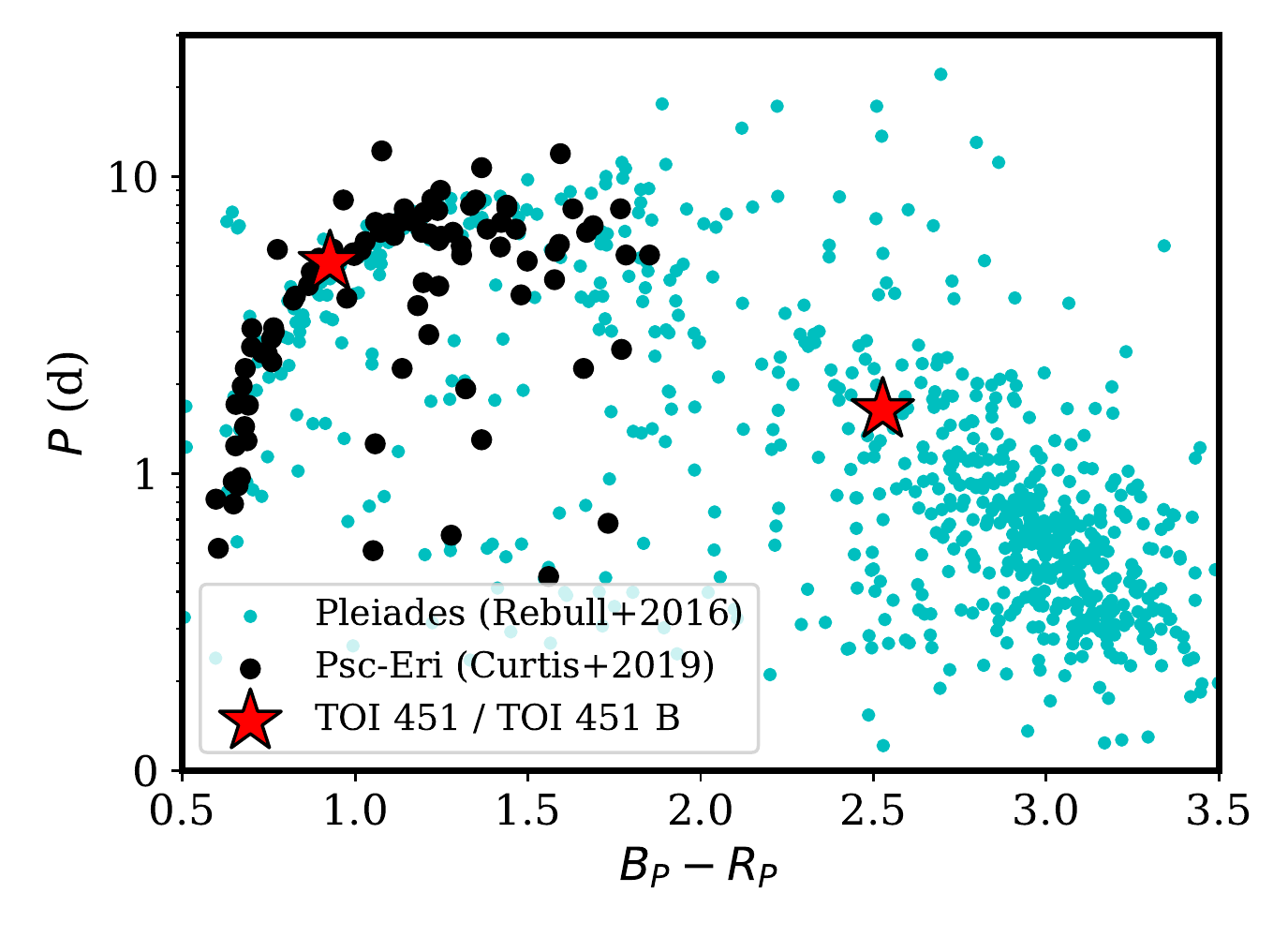}
    \caption{Rotation period as a function of color for TOI 451 and TOI 451 B (red stars), Psc-Eri members \citep[black circles;][]{CurtisTESS2019}, and Pleiades members \citep[cyan circles;][]{RebullRotation2016a}. Pleiades members are used to supplement the Psc-Eri members, which do not extend to later spectral types. Both TOI 451 and TOI 451 B have periods consistent with the color--rotation sequence that describes a $\sim120$ Myr old cluster.}
    \label{fig:rot}
\end{figure}

\begin{figure*}
    \centering
    \includegraphics[width=\textwidth]{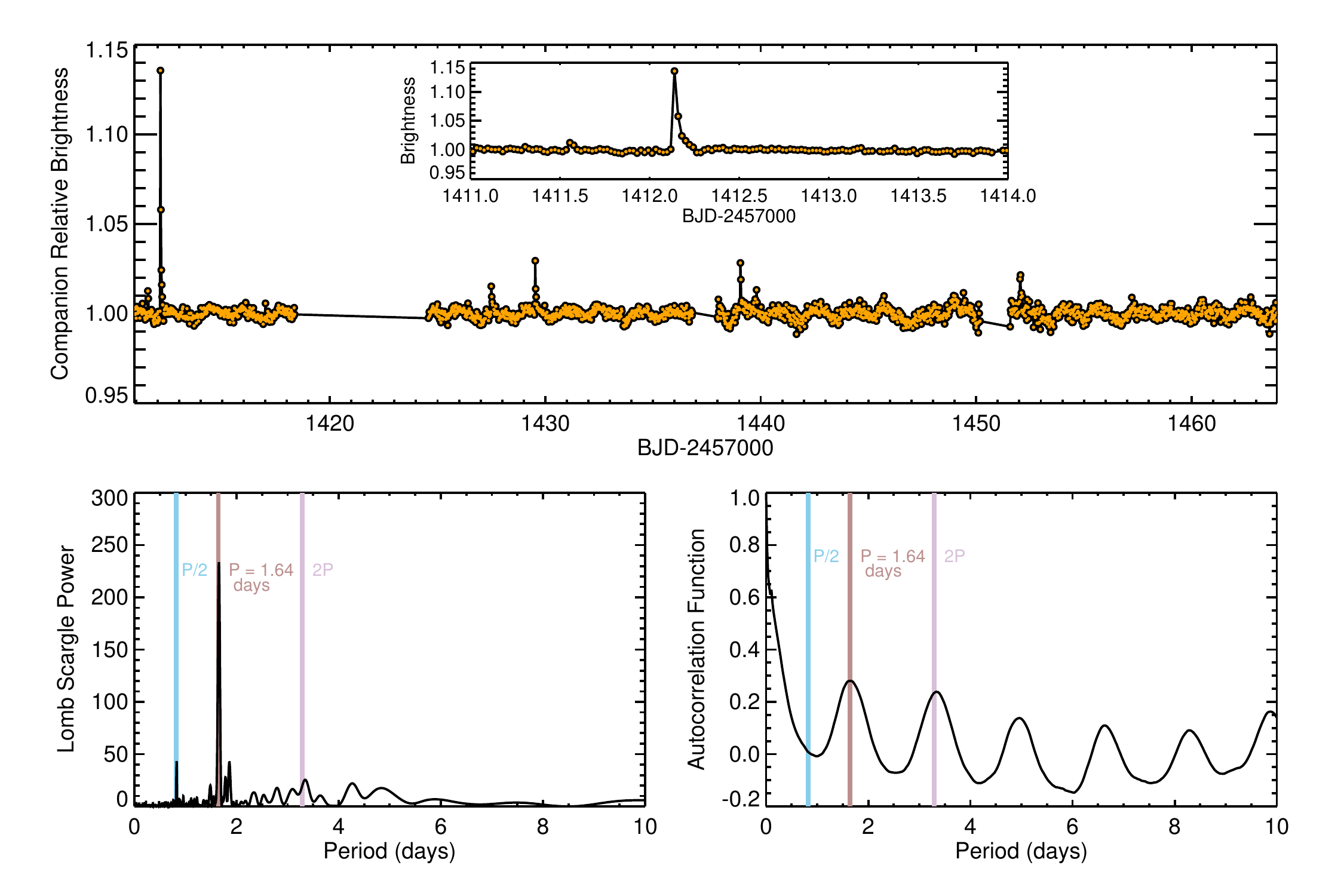}
    \caption{\textit{Top:} \TESS\ light curve of TOI 451 B. The inset shows additional detail of a large flare detected in the light curve. Typical error bars are about 0.002, smaller than the data points on the scale shown here. \textit{Bottom:} Lomb Scargle periodogram (\textit{Left}) and autocorrelation function (\textit{right}) of the \TESS\ light curve. Both the periodogram and autocorrelation function show a clear signal at 1.64 days. The true amplitude of the flares and signals in this light curve are larger than shown here due to diluting flux from TOI 451.   }
    \label{fig:companionrotation}
\end{figure*}

\subsection{Rotation period of TOI 451}

As described in \S\ref{sec:rotation}, we measured a rotation period of $5.1\pm0.1$ d, consistent with the 5.02 days reported in \citet{CurtisTESS2019}. As discussed in that work, this places TOI 451 on the slow sequence, the rotation--color sequence to which the initial distribution of periods converges for a uniform-age population. Figure \ref{fig:rot} places TOI 451 in the context of other members of Psc-Eri and Pleiades cluster members.

\subsection{Rotation period of TOI 451 B}

We extracted a light curve of TOI 451 B from the {\tess} 30 minute full frame images (the light curve would be from the composite object if TOI 451 B is itself a binary). TOI 451 and its companion(s) are only separated by 37 arcseconds, or about two TESS pixels, so the images of these two stars overlap substantially on the detector. The light curve of the companion TOI 451 B is clearly contaminated by the 14x brighter primary star. We therefore took a non-standard approach to extracting a light curve for the companion. We started with the flux time series of the single pixel closest to the position of TOI 451 B during the TESS observations, clipped out exposures with flags indicating low-quality datapoints and between times $1419 < BJD-2457000 < 1424$ (when a heater onboard the spacecraft was activated), and divided by the median flux value to normalize the light curve. We removed systematics and contaminating signals from TOI 451 by decorrelating the single-pixel light curve of TOI 451 B with mean and standard deviation quaternion time series \citep[see][]{Vanderburg2019}, a fourth order polynomial, and the flux from the pixel centered on TOI 451 (to model and remove any signals from the primary star). The resulting light curve, shown in Figure \ref{fig:companionrotation} shows several flares and a clear rotation signal. Before calculating rotation period metrics, we clipped out flares and removed points with times $1449 < BJD-2457000 < 1454$, which showed some residuals systematic effects. 

We measured the companion's rotation period by calculating the Lomb-Scargle periodogram and autocorrelation function from the two-sector TESS light curve. The ACF and Lomb-Scargle periodogram both showed a clear detection of a 1.64 day rotation period, though we cannot completely rule out the possibility that TOI 451 B's rotation period is actually an integer multiple of this period. 

We note that \citet{RebullRotation2016a}, in their analysis of the Pleiades, detect periods for 92\% of the members, and suggest the remaining non-detections are due to non-astrophysical effects. We have suggested TOI 451 B is a binary, which we might expect to manifest as two periodicities in the lightcurve. We only detect one period in our lightcurve; however, a second signal could have been impacted by systematics removal or be present at smaller amplitude than the 1.64 day signal, and so we do not interpret the lack of a second period further.

At around 100 Myr, stars of this type (early to mid M dwarfs) are in the midst of converging to the slow sequence. They may have a range of rotation periods, but are generally rotating with periods of a few days \citep{RebullRotation2016a}. While the Psc-Eri members studied in \citet{CurtisTESS2019} do not extend to stars as low mass as TOI 451 B, we can consider the similarly-aged Pleiades members as a proxy. Figure \ref{fig:rot} demonstrates that the $1.64$ day rotation period of TOI 451 B is typical for stars of its color at $120$~Myr.

\subsection{Age Diagnostics from \galex NUV Fluxes}

Chromospheric and coronal activity depend on stellar rotation and are thus also an age indicator \citep[e.g.][]{SkumanichTime1972}. We use excess UV emission as an age diagnostic for TOI 451 and TOI 451 B, considering the flux ratio $F_{NUV}/F_J$ as a function of spectral type. The use of this ratio for this purpose was suggested by \citet{ShkolnikSearching2011}.  More details on this general technique and applications to the Psc-Eri stream are provided in Appendix \ref{appendix}.

TOI 451 and TOI 451 B both were detected by \galex{} during short NUV exposures taken for the All-Sky Imaging Survey (AIS, \citealt{BianchiRevised2017}). Both stars were also observed with longer NUV exposures for the Medium-depth Imaging Survey (MIS), but only the primary's brightness was reported ($m_{nuv,p} = 16.674 \pm 0.006$ mag). Using the MIS data, we measured the secondary's brightness to be $m_{nuv,s} = 21.33 \pm 0.06$ mag (see Appendix \ref{appendix} for details).

In Figure~\ref{fig:galex}, we plot $F_{NUV}/F_J$ versus spectral type for  TOI 451 and TOI 451 B. Since both spectral type and color should be unaffected by a near-equal mass unresolved companion, the fact that TOI 451 B is a binary is not expected to impact this analysis. We also show isochronal sequences for the other members of Psc-Eri and the similarly-aged Pleiades cluster, using the latter to define the M dwarf regime at this age. To illustrate the expected fluxes for older stars, we show the sequence for the Hyades cluster. Details on the derivations of these isochronal sequences are given in Appendix \ref{appendix}. 

While solar-type stars only have excess NUV and X-ray fluxes for a short time, the NUV flux of early M dwarfs remains saturated to ages $\la 300$ Myr before sharply declining \citep{ShkolnikHAZMAT2014}. 
TOI 451 is consistent with the sequences for other G-dwarfs in all three clusters: as expected its UV excess is not a highly discriminating age diagnostic. However, the wide companion TOI 451 B sits above the Hyades sequence, and shows an NUV flux excess that is broadly consistent with the Pleiades sequence that is similar in age to Psc--Eri. This supports the membership of the TOI 451 system to the Psc-Eri stream.

\begin{figure}
    \centering
    \includegraphics[width=0.49\textwidth]{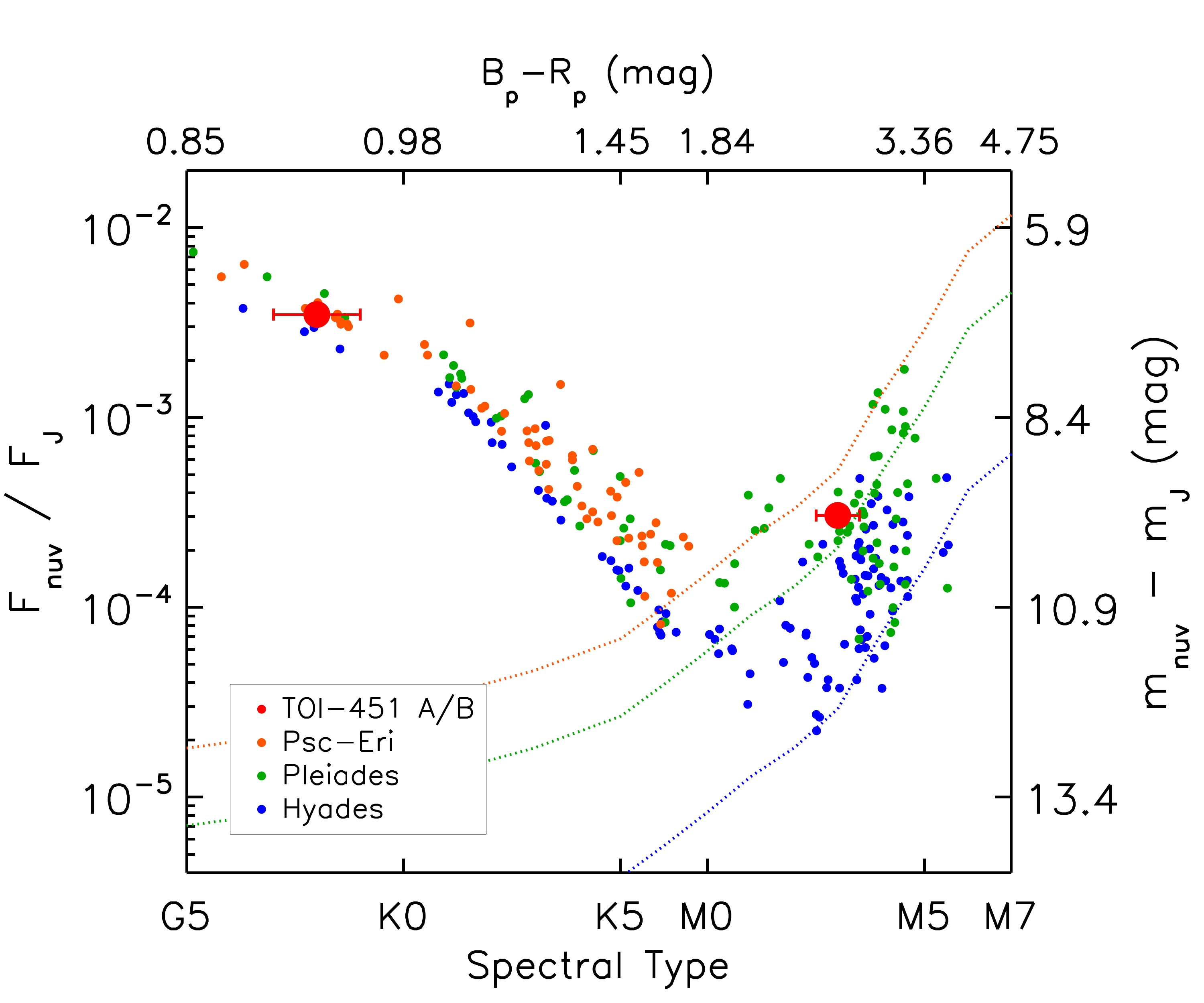}
    \caption{ GALEX NUV/NIR flux ratio ($F_{NUV}/F_J$ as a function of spectral type for TOI 451 and TOI 451 B, as well as several stellar populations spanning the age range where NUV fluxes are diagnostic of age. To allow direct comparison to other definitions of this youth diagnostic, we also show the $m_{nuv} - m_J$ and $B_p-R_p$ colors. 
    The dashed lines show the detection limits for each cluster shown; objects lying to the lower right would not have been detected in GALEX AIS at the assumed distance to that cluster (targeted pointings, e.g. for many Pleiades targets, extend deeper).}
    \label{fig:galex}
\end{figure}

\section{Analysis}\label{sec:analysis}

\subsection{Transit model}\label{sec:modeling}

We modeled the transit with {\tt misttborn} \citep{MannZodiacal2016a, JohnsonK22602018}. This routine uses {\tt batman} \citep{KreidbergBatman2015} to produce the transit model of \citet{MandelAnalytic2002} and explores the posterior with the Markov Chain Monte Carlo sampler {\tt emcee} \citep{Foreman-MackeyEmcee2013}.

To model the transits of each of the three planets, we fitted for the planet-to-star radius ratio $R_P/R_*$, impact parameter $b$, period $P$, and the epoch of the transit midpoint $T_0$. We assumed $R_P/R_*$ is the same in all filters. We also fitted for the mean stellar density ($\rho_*/\rho_\odot$). We used a quadratic limb-darkening law described by $g_{1,f}$ and $g_{2,f}$ for each filter $f$. This is a reasonable choice given that the host star is Sun-like \citep{EspinozaLimb2016}. We fitted the limb darkening parameters using the \cite{KippingEfficient2013} parameterization ($q_{1,f}$, $q_{2,f}$). For our first fit, we fixed the eccentricity $e$ to $0$. For our second fit, we allowed $e$ and the argument of periastron $\omega$ to vary, using the parameterization $\sqrt{e}\sin{\omega}$ and $\sqrt{e}\cos{\omega}$ \citep{FordImproving2006, EastmanEXOFAST2013}.

We used uniform priors for the planetary parameters, and a Gaussian prior on $\rho_*/\rho_\odot$ centered at 1.34 (Table \ref{tab:params}) with a $1\sigma$ width of 0.5. We placed Gaussian priors on the limb-darkening parameters based on theoretical values for a star with the temperature and radius given in Table \ref{tab:params}. We assume solar metallicity based on the mean iron abundance of the Psc-Eri members studied by \citet{HawkinsChemical2020}, who found $\rm{[Fe/H]}=-0.03$ dex with a dispersion of $0.04$-$0.07$ dex. For the Gaussian means of the limb darkening priors, we use the coefficients calculated with the {\tt Limb Darkening Toolkit} \citep[][; {\tt LDTK}]{ParviainenLDTK2015}, using the filter transmission curves available for LCO  $z_s$ \footnote{\url{https://lco.global/observatory/instruments/filters/}} and TESS\footnote{\url{https://heasarc.gsfc.nasa.gov/docs/tess/the-tess-space-telescope.html\#bandpass}}. The PEST bandpass is similar to that of MEarth, and following \citet{DittmannTemperate2017}, we adopted the filter profile from \citet{DittmannCalibration2016}. For {\it Spitzer}, we used the tabulated results from \citet{ClaretGravity2011}.
Due to systematic uncertainties in limb-darkening parameters \citep[]{MullerHighprecision2013, EspinozaLimb2015}, the potential impact of spots \citep{CsizmadiaEffect2013} and differences between LDTK and Claret coefficients, we used 0.1 for the $1\sigma$ width of the priors on $q$. 

For stellar rotation, we used the Gaussian process model of a mixture of simple harmonic oscillators introduced in \S\ref{sec:rotation}, with the parameters sampled in log-space (in the following, logarithms are all natural logarithms). The model includes the power at the rotation period $P_*$ (the primary signal) and at $P_*/2$ (the secondary signal). We fitted for the period of the primary signal $\ln{P_*}$, the amplitude of the primary signal $\ln{A_{1}}$, the relative amplitudes of the primary and secondary signals $m$ (where $A_1/A_2 = 1 + e^{-m}$), the decay timescale (or ``quality factor'') of the secondary signal $\ln{Q_{2}}$, and the difference in quality factors $\ln{\Delta Q}$ (where $\ln{\Delta Q}= \ln\left(Q_{1}-Q_{2}\right)$). We additionally included a photometric jitter term, $\sigma_\mathrm{GP}$.

We placed a Gaussian prior on $\ln{P_{\rm *}}$ centered at $1.6487$, based on the period from the WASP data, with a width of $0.05$. We use log-normal priors on the remaining parameters. We require $\ln{\Delta Q}>0$ to ensure that the primary signal has higher quality than the secondary, and $\ln{Q_2}>\ln{0.5}$ since we are modeling a signal with periodic behavior (cf. Equation 23 and Figure 1 in \citealt{Foreman-MackeyFast2017}).

The GP model was applied to data from \tess, LCO, and PEST, but not data from \spitzer. The latter was assumed to have no out-of-transit variability remaining after the corrections described in \S\ref{sec:data}. 

The autocorrelation length of our $e=0$ fit was 470 steps, and for our variable-eccentricity fit was 1200 steps. We ran the MCMC chain for 100 times the autocorrelation times with 100 walkers, discarding the first half as burn-in. 
Figures \ref{fig:tess} and \ref{fig:other} shows the best-fitting models overlain on the data, and Table \ref{tab:results} lists the median and 68\% confidence limits of the fitted planetary and stellar parameters.

\begin{deluxetable*}{l | r r r | r r r}
\tabletypesize{\tiny}
\setlength{\tabcolsep}{2pt}
\renewcommand{\arraystretch}{1.5}
\tablecaption{Transit fitting results for the TOI 451 system \label{tab:results}}
\tablewidth{0pt}
\tablehead{\colhead{Parameter} & \colhead{Planet b} &  \colhead{Planet c} &  \colhead{Planet d} & \colhead{Planet b} &   \colhead{Planet c} &   \colhead{Planet d}\\ 
\colhead{} &  \multicolumn{3}{c}{$e$,$\omega$ fixed} &\multicolumn{3}{c}{$e$,$\omega$ free}
 }
\startdata
\multicolumn{7}{c}{Fitted Transit Parameters} \\
\hline
$T_0$ (BJD) & $1410.9900^{+0.0056}_{-0.0037}$ & $1411.7956^{+0.0048}_{-0.0026}$ & $1416.63478^{+0.00088}_{-0.00092}$ & $1410.9909^{+0.0046}_{-0.0042}$ & $1411.7961^{+0.0039}_{-0.0030}$ & $1416.63499^{+0.00097}_{-0.00093}$ \\ 
$P$ (days) & $1.858703^{+2.5\times10^{-5}}_{-3.5\times10^{-5}}$  & $9.192522^{+6.0\times10^{-5}}_{-10\times10^{-5}}$  & $16.364988 \pm 4.4\times10^{-5}$  & $1.858701^{+2.7\times10^{-5}}_{-3.3\times10^{-5}}$  & $9.192523^{+6.4\times10^{-5}}_{-8.4\times10^{-5}}$  & $16.364981^{+4.7\times10^{-5}}_{-4.9\times10^{-5}}$ \\ 
$R_P/R_{\star}$ & $0.0199^{+0.0010}_{-0.0011}$  & $0.03237^{+0.00065}_{-0.00070}$  & $0.04246 \pm 0.00044$  & $0.0203^{+0.0014}_{-0.0011}$  & $0.03206^{+0.00090}_{-0.00085}$  & $0.04205^{+0.00050}_{-0.00045}$ \\ 
$b$ & $0.22^{+0.20}_{-0.15}$ & $0.139^{+0.121}_{-0.096}$ & $0.387^{+0.047}_{-0.039}$ & $0.42^{+0.32}_{-0.28}$ & $0.52^{+0.23}_{-0.32}$ & $0.23^{+0.18}_{-0.16}$ \\ 
$\rho_{\star}$ ($\rho_{\odot}$) & \multicolumn{3}{c|}{$1.294^{+0.061}_{-0.088}$}  & \multicolumn{3}{c}{$1.41^{+0.15}_{-0.16}$} \\ 
$q_{1,1}$ & \multicolumn{3}{c|}{$0.344^{+0.080}_{-0.078}$}  & \multicolumn{3}{c}{$0.385^{+0.083}_{-0.084}$} \\ 
$q_{2,1}$ & \multicolumn{3}{c|}{$0.394^{+0.068}_{-0.082}$}  & \multicolumn{3}{c}{$0.391^{+0.070}_{-0.088}$} \\ 
$q_{1,2}$ & \multicolumn{3}{c|}{$0.325^{+0.095}_{-0.096}$}  & \multicolumn{3}{c}{$0.325^{+0.099}_{-0.086}$} \\ 
$q_{2,2}$ & \multicolumn{3}{c|}{$0.365^{+0.080}_{-0.096}$}  & \multicolumn{3}{c}{$0.364^{+0.082}_{-0.100}$} \\ 
$q_{1,3}$ & \multicolumn{3}{c|}{$0.058^{+0.076}_{-0.045}$}  & \multicolumn{3}{c}{$0.058^{+0.080}_{-0.045}$} \\ 
$q_{2,3}$ & \multicolumn{3}{c|}{$0.341^{+0.085}_{-0.094}$}  & \multicolumn{3}{c}{$0.324^{+0.089}_{-0.092}$} \\ 
$q_{1,4}$ & \multicolumn{3}{c|}{$0.036^{+0.042}_{-0.025}$}  & \multicolumn{3}{c}{$0.027^{+0.038}_{-0.020}$} \\ 
$q_{2,4}$ & \multicolumn{3}{c|}{$0.155^{+0.094}_{-0.086}$}  & \multicolumn{3}{c}{$0.159^{+0.095}_{-0.090}$} \\ 
$\sqrt{e}\sin\omega$ & \multicolumn{3}{c|}{$\cdots$} & $-0.23^{+0.19}_{-0.21}$ & $-0.26^{+0.16}_{-0.19}$ & $0.02^{+0.11}_{-0.13}$ \\ 
$\sqrt{e}\cos\omega$ & \multicolumn{3}{c|}{$\cdots$} & $-0.09^{+0.42}_{-0.38}$ & $0.01^{+0.36}_{-0.43}$ & $-0.01^{+0.32}_{-0.29}$ \\ 
\hline
\multicolumn{7}{c}{Fitted Gaussian Process Parameters} \\
\hline
$\log{P_{\mathrm{GP}}}$ (day) & \multicolumn{3}{c|}{$1.636^{+0.027}_{-0.024}$} & \multicolumn{3}{c}{$1.637^{+0.027}_{-0.024}$} \\ 
$\log{A_{\mathrm{GP}}}$ (\%$^2$) & \multicolumn{3}{c|}{$-12.08^{+0.36}_{-0.26}$} & \multicolumn{3}{c}{$-12.1^{+0.32}_{-0.25}$} \\ 
$\log{Q1_{\mathrm{GP}}}$ & \multicolumn{3}{c|}{$1.62^{+0.93}_{-0.94}$} & \multicolumn{3}{c}{$1.59^{+0.89}_{-0.95}$} \\ 
$\log{Q2_{\mathrm{GP}}}$ & \multicolumn{3}{c|}{$1.25^{+0.33}_{-0.29}$} & \multicolumn{3}{c}{$1.25^{+0.34}_{-0.30}$} \\
Mix$_\mathrm{Q1,Q2}$ & \multicolumn{3}{c|}{$4.6 \pm 3.5$} & \multicolumn{3}{c}{$4.8^{+3.6}_{-3.2}$} \\ 
$\sigma_{\mathrm{GP}}$ (\%) & \multicolumn{3}{c|}{$-15.0^{+3.1}_{-3.4}$} & \multicolumn{3}{c}{$-15.0^{+3.1}_{-3.3}$} \\ 
\hline
\multicolumn{7}{c}{Derived Transit Parameters} \\
\hline
$a/R_{\star}$ & $6.93^{+0.11}_{-0.16}$  & $20.12^{+0.31}_{-0.47}$  & $29.56^{+0.46}_{-0.69}$  & $6.63^{+0.51}_{-0.89}$  & $18.9^{+1.4}_{-2.2}$  & $30.69^{+0.82}_{-1.0}$ \\ 
$i$ ($^{\circ}$) & $88.2^{+1.2}_{-1.7}$  & $89.61^{+0.27}_{-0.36}$  & $89.25^{+0.084}_{-0.1}$  & $86.5^{+2.3}_{-2.9}$  & $88.49^{+0.95}_{-0.67}$  & $89.56^{+0.31}_{-0.35}$ \\ 
$\delta$ (\%) & $0.0396^{+0.0041}_{-0.0042}$  & $0.1048^{+0.0043}_{-0.0045}$  & $0.1803 \pm 0.0037$  & $0.0412^{+0.0059}_{-0.0043}$  & $0.1028^{+0.0059}_{-0.0054}$  & $0.1768^{+0.0042}_{-0.0038}$ \\ 
$T_{14}$ (days) & $0.0849^{+0.0024}_{-0.0054}$ & $0.1483 \pm 0.0016$ & $0.1707 \pm 0.001$ & $0.082^{+0.02}_{-0.017}$ & $0.137^{+0.028}_{-0.031}$ & $0.171^{+0.018}_{-0.017}$ \\ 
$T_{23}$ (days) & $0.0815^{+0.0024}_{-0.0058}$& $0.1387^{+0.0015}_{-0.0016}$& $0.1543 \pm 0.0011$& $0.078^{+0.019}_{-0.018}$& $0.125^{+0.026}_{-0.032}$& $0.156^{+0.016}_{-0.017}$ \\ 
FPP parameter & $0.119^{+0.308}_{-0.094}$ & $0.044^{+0.11}_{-0.038}$ & $0.354^{+0.101}_{-0.072}$ & $0.44^{+1.32}_{-0.38}$ & $0.7^{+1.12}_{-0.61}$ & $0.12^{+0.28}_{-0.11}$ \\ 
$T_{\mathrm{peri}}$ (BJD) & $1410.99^{+0.0056}_{-0.0037}$  & $1411.7956^{+0.0048}_{-0.0026}$  & $1416.63478^{+0.00088}_{-0.00092}$  & $1411.26^{+0.39}_{-0.89}$  & $1410.4^{+4.6}_{-2.1}$  & $1416.8^{+3.3}_{-3.8}$ \\ 
$g_{1,1}$ & \multicolumn{3}{c|}{$0.452^{+0.093}_{-0.100}$}  & \multicolumn{3}{c}{$0.479^{+0.094}_{-0.100}$} \\ 
$g_{2,1}$ & \multicolumn{3}{c|}{$0.122^{+0.100}_{-0.078}$} & \multicolumn{3}{c}{$0.132^{+0.113}_{-0.086}$} \\ 
$g_{1,2}$ & \multicolumn{3}{c|}{$0.40 \pm 0.12$}& \multicolumn{3}{c}{$0.41^{+0.12}_{-0.13}$} \\ 
$g_{2,2}$ & \multicolumn{3}{c|}{$0.148^{+0.116}_{-0.089}$} & \multicolumn{3}{c}{$0.154^{+0.111}_{-0.095}$} \\ 
$g_{1,3}$ & \multicolumn{3}{c|}{$0.151^{+0.093}_{-0.080}$} & \multicolumn{3}{c}{$0.145^{+0.090}_{-0.077}$} \\ 
$g_{2,3}$ & \multicolumn{3}{c|}{$0.067^{+0.082}_{-0.047}$} & \multicolumn{3}{c}{$0.075^{+0.083}_{-0.050}$} \\ 
$g_{1,4}$ & \multicolumn{3}{c|}{$0.052^{+0.048}_{-0.033}$} & \multicolumn{3}{c}{$0.046^{+0.045}_{-0.029}$} \\ 
$g_{2,4}$ & \multicolumn{3}{c|}{$0.122^{+0.081}_{-0.061}$} & \multicolumn{3}{c}{$0.106^{+0.080}_{-0.057}$} \\ 
$R_P$ ($R_{\oplus}$) & $1.91 \pm 0.12$ & $3.1 \pm 0.13$ & $4.07 \pm 0.15$ & $1.94^{+0.15}_{-0.13}$ & $3.07 \pm 0.14$ & $4.03 \pm 0.15$ \\ 
$a$ (AU) & $0.0283^{+0.0011}_{-0.0012}$ & $0.0823^{+0.0033}_{-0.0036}$ & $0.1208^{+0.0048}_{-0.0052}$ & $0.0271^{+0.0023}_{-0.0038}$ & $0.0771^{+0.0066}_{-0.0093}$ & $0.1255^{+0.0057}_{-0.0065}$ \\ 
$T_{\mathrm{eq}}$ (K) & $1491^{+23}_{-19}$  & $875^{+13}_{-11}$  & $722^{+11}_{-9}$  & $1524^{+100}_{-60}$  & $903^{+53}_{-36}$  & $708^{+15}_{-12}$ \\ 
$e$ & \multicolumn{3}{c|}{$\cdots$} & $0.19^{+0.20}_{-0.14}$ & $0.2^{+0.18}_{-0.14}$ & $0.057^{+0.133}_{-0.040}$ \\ 
$\omega$ ($^{\circ}$) & \multicolumn{3}{c|}{$\cdots$} & $238^{+85}_{-48}$ & $266 \pm 63$ & $170^{+170}_{-120}$ \\ 
\enddata
\tablecomments{$T_{\mathrm{eq}}$ assumes an albedo of 0 (the planets reflect no light) with no uncertainty. The planetary radii listed above are from the joint fit of {\it TESS}, ground-based, and {\it Spitzer} data. For the {\it TESS}-only fit, $R_P/R_{\star}$ for planets b, c and d, respectively are: $0.0194 \pm 0.0011$, $0.0286 \pm 0.0016$, and $0.0418^{+0.0012}_{-0.0013}$. The largest difference is for c. For the {\it Spitzer}-only fit of c, $R_P/R_{\star}=0.034 \pm 0.001$. We suggest adopting the $e=0$ fit since the eccentricities from the variable-eccentricity fit are consistent with 0.}
\end{deluxetable*}

\subsection{Additional investigations into the transit parameters}

We performed fits of the \tess\ data using {\tt EXOFASTv2}\footnote{\url{https://github.com/jdeast/EXOFASTv2}} \citep{2019arXiv190709480E} software package, which simultaneously fits the transits with the stellar parameters. We first removed the stellar variability using the GP model described in Section \ref{sec:rotation}, masking the transit prior to fitting. The stellar parameters were constrained by the spectral energy distribution (SED) and MESA Isochrones and Stellar Tracks (MIST) stellar evolution models \cite{DotterMESA2016, ChoiMESA2016a}.
We enforced Gaussian priors on $T_\mathrm{eff}$, [Fe/H], and \logg following Section \ref{sec:measure}, and on the Gaia DR2 parallax corrected for a systematic offset \citep{StassunEvidence2018}. We used a Gaussian prior on age of $125\pm15$ Myr, and restricted extinction to $<0.039$ \citep{2011ApJ...737..103S}. The stellar parameters from {\tt EXOFASTv2} are: $\teff=5563 \pm 44 K$, $M_*=0.94\pm0.025$ and $R_*=0.834\pm0.01$.
We generally found excellent agreement with the best-fitting planetary parameters from {\tt misttborn}.

We also fitted the \tess\ and \spitzer\ data sets independently. There is a $2\sigma$ discrepancy between the \spitzer\ transit depths of TOI 451 c ($R_P/R_{\star} = 0.034 \pm 0.001$) and the \tess\ transit depth ($R_P/R_{\star} = 0.029 \pm 0.002$). No significant differences are seen in the transit depths of TOI 451 d, which we might expect if there was dilution in the \tess\ data. The mildly larger planetary radius for TOI 451 c measured at $4.5\micron$ relative to that measured at {\it TESS}'s red-optical bandpass could be due to occulted starspots\footnote{A planet transiting over starspots would block a smaller fraction of the star's light than it would otherwise and would therefore appear to have a smaller planetary radius. Since the contrast between spots and the photosphere is larger in the optical than the infrared, this effect would be stronger in the {\it TESS} data compared to the \spitzer\ data.}, unocculted plages, or atmospheric features. Given that the transits of c and d were observed at similar times, we might also expect differences in the depths of TOI 451 d if starspots caused the discrepancy for c; however, the nature of starspots (size, temperature, location/active latitudes, longevity) is not well-understood. Alternatively, the difference in transit depths for TOI 451 c could derive from its atmosphere. It is likely, however, that difference arises from systematics in the \spitzer\ data, as discussed in \S\ref{sec:spitzer}.

\subsection{False positive analysis}

We considered four different false-positive scenarios for each of the candidate planets: 1) there is uncorrected stellar variability, 2) TOI 451 is an eclipsing binary, 3) TOI 451 is a hierarchical eclipsing system, 4) there is a background or foreground eclipsing system. Relevant to the blend scenarios, transits are visible in the {\it Spitzer} data even when shrinking or shifting the aperture, indicating that the signal lands within 2 {\it Spitzer} pixels (2.4\arcsec) of \target. Speckle imaging and \gaia\ RUWE rule out companions brighter than these limits (i.e., bright enough to cause the transits) down to $0.2\arcsec$, so a very close blend is required.

We ruled out stellar variability for all three planets from the {\it Spitzer} transits. For any spot contrast, stellar variation due to rotation and spots/plages will always be weaker at {\it Spitzer} wavelengths compared to {\it TESS}. For \target, out-of-transit data taken by {\it Spitzer} shows a factor of $\simeq$4 lower variability than the equivalent baseline in {\it TESS} data. Thus, if any transit was due purely to uncorrected stellar variation, the shape, duration, and depth would be significantly different or the entire transit would not present in the {\it Spitzer} photometry. 

We used the source brightness parameter from \citet{Vanderburg2019} to constrain the magnitude a putative blended source (bound or otherwise). The parameter, $\Delta m$, relates the ingress or egress duration to transit duration and reflects the true radius ratio, independent of whether there is contaminating flux: $\Delta m \leq 2.5\log_{10}(T^2_{12}/T^2_{13}/\delta)$. Here, $\delta$ is the transit depth, $T_{12}$ is the ingress/egress duration and $T_{13}$ the time between the first and third contact. We calculated $\Delta m$ for the posterior samples for our variable-eccentricity transit fit, and took the 99.7\% confidence limit. We find $\Delta m <4.6$, $2.0$, $1.0$ for TOI 451 b, c and d, respectively. 

Using the brightness constraints from our imaging data and the $\Delta m$ parameter, we statistically rejected the scenario where any of the transits are due to an unassociated field star, either an eclipsing binary or an unassociated transiting planetary system \citep{2011ApJ...738..170M}. We drew information on every star within 1 degree of \target{} from {\it Gaia} DR2 satisfying the most conservative brightness limits ($\Delta T <4.6$). This yielded a source density of $\simeq$3500 stars per square degree, suggesting a negligible $\simeq7\times10^{-5}$ field stars that were missed by our speckle data and still bright enough to produce the candidate transit associated with b. Constraints are stronger for the other planets. 

To investigate scenarios including a bound companion blended with the source, we first considered the constraint placed by the multi-wavelength transit depths. As explained in \citet{Desert2015}, if the transit signals were associated with another star in the aperture, the transit depth observed by {\it Spitzer} would be deeper than that observed by {\it TESS}, owing to the decreased contrast ratio between the target and the blended star (which must be fainter as is assumed to be cooler) at 4.5\um\ compared to 0.75\um. Thus, the ratio of the {\it Spitzer}-to-{\it TESS} transit depths ($\delta S/\delta T$) provides a range of possible {\it TESS}-{\it Spitzer} colors of the putative companion ($C_{TS,\rm{comp}}$) in terms of the combined (unresolved) color ($C_{TS,\rm{combined}}$). Following Tofflemire et al. (in prep), we used the 95-percentile range for $\delta S/\delta T$ for each transit to derive the putative companion color: 
\begin{equation}
C_{TS,\rm{comp}}<C_{TS,\rm{combined}} + 2.5 \log_{10}(\frac{\delta S}{\delta T}).
\end{equation}
Adopting the weakest constraints from the three planets gives a $C_{TS,\rm{comp}}$ between 1.09 and 1.30 magnitudes, indicating that the putative bound companion must be similar in color to the target star. 

\begin{figure}
    \centering
    \includegraphics[width=0.47\textwidth]{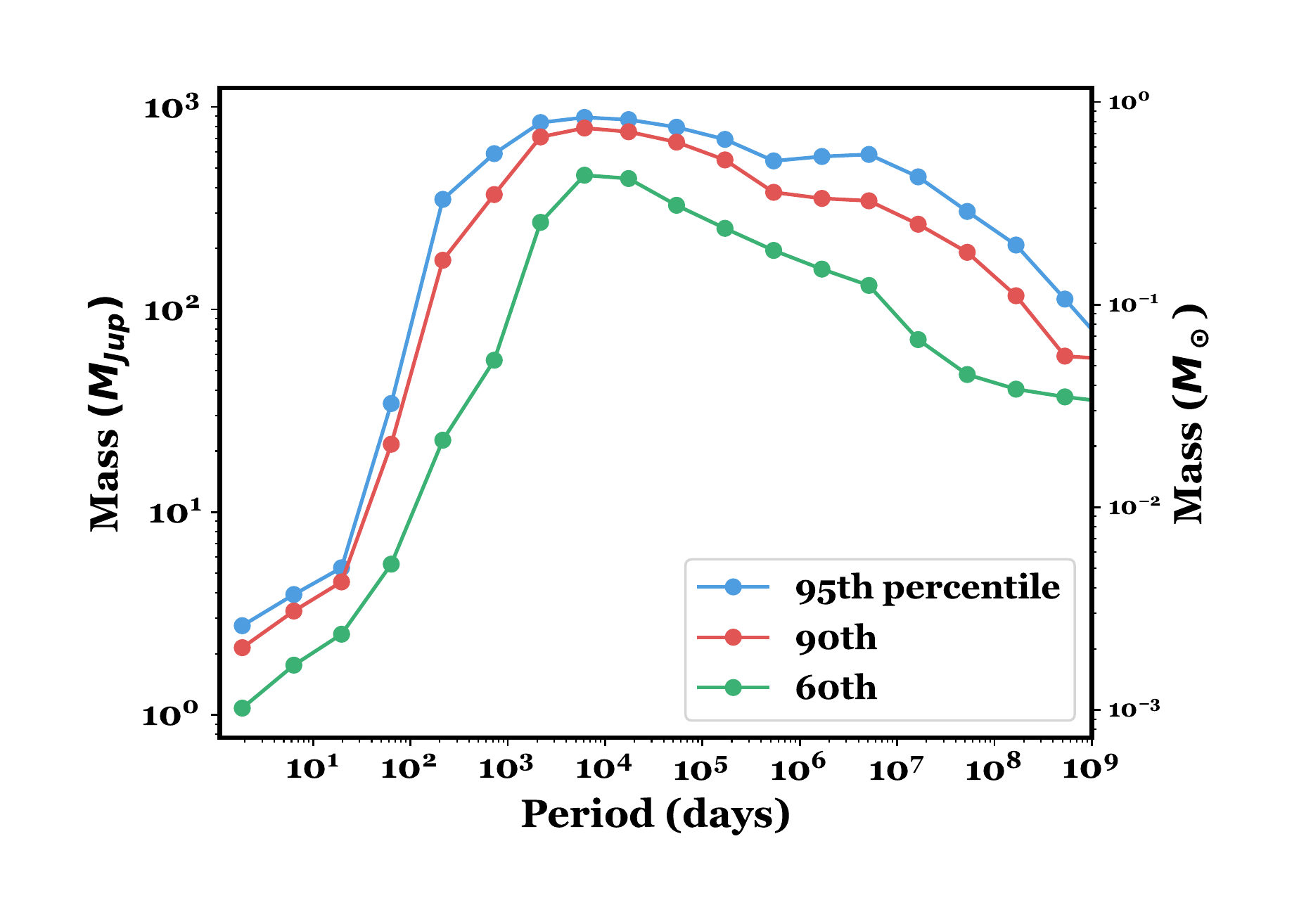}
    \caption{Distribution of surviving binaries from our false-positive simulation. In this case, the companion is forced to eclipse the star, as would be required to produce a transit-like signal. The curve rules out all stellar and brown dwarf companions at periods matching the three planets. }
    \label{fig:EBs}
\end{figure}

We then simulated binary systems following Wood et al. (in prep) which we compared to the observational data and constraints. In short, we generated 5 million binaries following the period and mass ratio from \citet{RaghavanSurvey2010} and the eccentricity distribution from \citet{Price-WhelanClose2020}. For each binary, we calculated the expected radial velocity curve, magnitude, and projected separation at the epoch of the speckle data. We then compared the generated models for a given simulated binary to our radial velocities, speckle data, and {\it Gaia} imaging and astrometry (Section~\ref{sec:comp}). To account for stellar variability, we added 50\,\mps\ error to the velocity measurements. Binaries were then rejected based on the probability of the observational constraints being consistent with the binary star parameters by chance (color, source brightness, radial velocities, and all imaging/photometric data). We ran two versions of this simulation, one where a single companion was forced to eclipse the primary (for ruling out eclipsing binaries), and one where the binary's orbital inclination was unrestricted (for hierarchical systems). The former set ruled out all non-planetary signals at periods matching any of the three planets (Figure~\ref{fig:EBs}). In the latter simulation, less than 0.01\% of the simulated blends passed all observational constraints for the signals associated with any of the three planets. 

We now take into consideration the probability that any given star happens to be an eclipsing binary (about 1\%; \citealt{2016AJ....151...68K}), compared to the probability of a transiting planet (also about 1\%; \citealt{2018ApJS..235...38T}). The chance alignment of unassociated field stars is therefore $7\times10^{-5}$ times as likely as the transiting planet scenario, and the hierarchical triple $1\times10^{-4}$ times as likely. The probability of the host star itself being an eclipsing binary is negligible. The transiting planet hypothesis is significantly more likely than the non-transiting planet hypothesis, with a false positive probability of about $2\times10^{-4}$. The probability of a star hosting a planet discovered by {\it TESS} is likely lower (about 0.05\% based on current search statistics; \citealt{GuerreroTESS2020}) than assumed here. On the other hand, studies of the \kepler\ multi-planet systems indicate that the false positive rate for systems with two or more transiting planets is low \citep{LissauerAlmost2012, LissauerValidation2014}, about a factor of 15 less for TESS planets \citep{GuerreroTESS2020}, which would more than compensate for our assumption. The false positive probability of $2\times10^{-4}$ is therefore an overestimate.

We have ruled out instrumental variability, chance alignment of unassociated field stars (including both eclipsing binaries and planetary systems), and bound eclipsing binaries or heirarchical triples, and therefore reject the false positives identified at the beginning of this section. We consider all three planets validated at high confidence. 

\section{Summary and discussion}\label{sec:summary}

TOI 451 b, c and d are hot planets in close orbits around a young, Sun-like star. The inner planet, TOI 451\,b, has a period of $1.9$ days and a radius of $1.9$\rearth. This places it within or below the radius valley as defined in \citet[][cf. their Figure 6]{VanEylenAsteroseismic2018}, although there are very few larger planets ($2-4$\rearth) at such short orbital periods. At $3.1$ and $4.1$\rearth, the outer two planets sit well above the radius valley. Since they are young and hot ($T_\mathrm{eq} = 720$ to $1500$ K assuming $0$ albedo) and could be low-mass, their observed radii may be impacted by high-altitude hazes \citep{GaoDeflating2019}.

At $120$ Myr, the solar-mass host star has completed the most magnetically active part of its lifetime and the era of strongest photoevaporative mass loss is expected to be complete \citep{OwenEvaporation2017}. However, \cite{RogersUnveiling2020}  showed that overall photoevaporatively-driven radius evolution of a synthetic population was not complete until around 1 Gyr. Core-powered mass loss \citep{GinzburgCorepowered2018} would be expected to shape planetary radii on timescales of 1 Gyr \citep{GuptaSignatures2020}.  Thus, these planets may still be undergoing observable atmospheric mass loss.


We estimated the planetary masses using the non-parameteric mass-radius relation from \citet{NingPredicting2018}, which is based on the full {\it Kepler} data set.\footnote{\url{https://github.com/shbhuk/mrexo}; \cite{KanodiaMassRadius2019}} This assumes these young planets obey the same mass-radius relation as older stars, which may be inaccurate. We found masses for b, c, and d of $5^{+7}_{-3}$\mearth, $7^{+9}_{-4}$\mearth, and $8^{+11}_{-4}$\mearth, respectively. The expected RV amplitudes are about $2$ m/s for the three planets.  Mass measurements of the planets are likely to be challenging due to the expected jitter and $5$ d stellar rotation period, but would provide valuable information on the planetary compositions.

Precise transit timing variations have the potential to yield measurement of the masses of the outer two planets. To estimate the TTV amplitudes, we assumed no other planets interact dynamically with the three we observe and used \texttt{TTV2Fast2Furious}\footnote{\url{https://github.com/shadden/TTV2Fast2Furious}} (\citealt{HaddenShadden2019}; see \citealt{HaddenProspects2019}). We estimated TTV amplitudes (measured peak to zero) of 2 min for TOI 451 c and d using our variable-eccentricity fit, or 30 sec for c and 1 min for d for our $e=0$ fit. The periodicity is about 75 d.
At present, CHEOPS \citep{2013EPJWC..4703005B} is the photometric facility that can provide the highest-precision photometry for this system. Due to its Sun-Synchronous orbit, CHEOPS can monitor TOI 451 with at least 50\% observation efficiency for approximately one month every year. We used the equations of \cite{2014ApJ...794...92P} to estimate the uncertainty on the mid-transit time achievable with CHEOPS for TOI 451 c and d from their respective ingress and total transit durations. We assumed  
an exposure time of 30 s. We estimated the precision in the mid-transit to be 2 and 1.6 min per transit for TOI 451 c and d, respectively. For the variable-eccentricity fit results, TTV measurements for these two planets may thus be within reach.

Using the system parameters reported in this paper and the estimated
planet masses listed above, along with the estimated uncertainties on
all parameters, we estimated the Transmission Spectroscopy Metric
\citep[TSM;][]{2018PASP..130k4401K} for the three TOI~451 planets. S/N scales linearly with TSM for stars with $J>9$ ($J=9.6$ mag for TOI 451), with a scale factor of around $1.2$. For
planets b, c, and d we find TSM $=36\pm22$, $59\pm35$, and $98\pm57$, respectively. These fairly large uncertainties result from the
as-yet unknown planet masses, but even accounting for the likely
distribution of masses we find 95.4\% (2$\sigma$) lower limits of
TSM$>14$, $26$, and $44$, respectively.  Planets d is likely to be among the best known planets in its
class for transmission spectroscopy \citep[cf.\ Table~11; ][]{2020AJ....159..239G}. 
 
In summary, we have validated a three-planet system around TOI 451, a solar mass star in the 120-Myr Pisces-Eridanis stream. The planets were identified in {\it TESS}, and confirmed to occur around the target star with some of the final observations taken by {\it Spitzer} as well as ground-based photometry and spectroscopy. We confirmed that TOI 451 is a member of Psc-Eri by considering the kinematics and lithium abundance of TOI 451, and the rotation periods and NUV activity levels of both TOI 451 and its wide-binary companion TOI 451 B (Appendix \ref{appendix} discusses the utility of NUV activity for identifying new low-mass members of Psc-Eri). The stellar rotation axis, the planetary orbits, and the binary orbit may all be aligned; and there is evidence for a debris disk. The synergy of all-sky, public datasets from {\it Gaia} and {\it TESS} first enabled the identification of this star as a young system with a well-constrained age, and then the discovery of its planets.

\facilities{LCO, SALT, TESS, SOAR, PEST, WASP, LCO, \spitzer}

\software{
{\tt numpy} \cite{harris2020array},
{\tt scipy} \cite{2020SciPy-NMeth},
{\tt matplotlib} \cite{Hunter:2007},
{\tt misttborn} \cite{MannZodiacal2016a, JohnsonK22602018},
{\tt celerite} \cite{Foreman-MackeyFast2017},
{\tt batman} \cite{KreidbergBatman2015},
{\tt LDTK} \cite{ParviainenLDTK2015},
{\tt corner} \cite{Foreman-MackeyCorner2016},
{\tt emcee} \cite{Foreman-MackeyEmcee2013},
{\tt pymc3} \cite{SalvatierProbabilistic2016},
{\tt exoplanet} \cite{exoplanet:exoplanet},
{\tt starry} \cite{exoplanet:luger18},
{\tt astroquery} \cite{2019AJ....157...98G},
{\tt astropy} \cite{2013A&A...558A..33A, TheAstropyCollaborationAstropy2018},
{\tt AstroImageJ} \cite{Collins:2017}
}

AWM was supported through NASA's Astrophysics Data Analysis Program (80NSSC19K0583). This material is based upon work supported by the National Science Foundation Graduate Research Fellowship Program under Grant No. DGE-1650116 to PCT. KH has been partially supported by a TDA/Scialog  (2018-2020) grant funded by the Research Corporation and a TDA/Scialog grant (2019-2021) funded by the Heising-Simons Foundation. KH also acknowledges support from the National Science Foundation grant AST-1907417 and from the Wootton Center for Astrophysical Plasma Properties funded under the United States Department of Energy collaborative agreement DE-NA0003843. D.D. acknowledges support from the TESS Guest Investigator Program grant 80NSSC19K1727 and NASA Exoplanet Research Program grant 18-2XRP18 2-0136. I.J.M.C. acknowledges support from the NSF through grant AST-1824644,
and from NASA through Caltech/JPL grant RSA-1610091.

This work makes use of observations from the Las Cumbres Observatory (LCO) network. Some of the observations reported in this paper were obtained with the Southern African Large Telescope (SALT).
This  work  has  made  use  of  data  from  the  EuropeanSpace Agency (ESA) mission {\it Gaia} (\url{https://www.cosmos.esa.int/gaia}), processed by the {\it Gaia} Data Processing and Analysis  Consortium  (DPAC, \url{https://www.cosmos.esa.int/web/gaia/dpac/consortium}).  Funding  for  the  DPAC has been provided by national institutions, in particular the institutions  participating  in  the {\it Gaia} Multilateral  Agreement.
Based in part on observations obtained at the Southern Astrophysical Research (SOAR) telescope, which is a joint project of the Ministério da Ciência, Tecnologia e Inovações do Brasil (MCTI/LNA), the US National Science Foundation’s NOIRLab, the University of North Carolina at Chapel Hill (UNC), and Michigan State University (MSU).

We acknowledge the use of public \tess\ Alert data from pipelines at the \tess\ Science Office and at the \tess\ Science Processing Operations Center. Resources supporting this work were provided by the NASA High-End Computing (HEC) Program through the NASA Advanced Supercomputing (NAS) Division at Ames Research Center for the production of the SPOC data products.

This paper includes data collected by the \tess\ mission, which are publicly available from the Mikulski Archive for Space Telescopes (MAST). Funding for the \tess\ mission is provided by NASA’s Science Mission directorate. This research has made use of the Exoplanet Follow-up Observation Program website, which is operated by the California Institute of Technology, under contract with the National Aeronautics and Space Administration under the Exoplanet Exploration Program.

\bibliography{PscEri.bib}

\appendix

\section{UV excess as a way to identify young stars in Psc--Eri}\label{appendix}

Chromospheric activity is indicative of stellar ages in a parallel to rotational periods. Fast-rotating stars are active stars, and this activity manifests in young stars via both spectral line emission (most notably Ca H\&K and H$\alpha$; \citealt{WilsonProbable1963, FeigelsonDiscovery1981}) and excess emission in the ultraviolet and X-rays \citep[e.g.][]{KuEinstein1979}. 
X-ray emission from the \rosat All-Sky Survey was used for searches of nearby young populations \citep[e.g.][]{WalterXray1994, 1996A&A...312..439W, TorresSearch2006}. The better sensitivity of the \galex ultraviolet space observatory opened up new opportunities \citep[e.g.][]{FindeisenUltravioletselected2010, ShkolnikSearching2011, RodriguezNew2011}. 

UV flux measurements allow age-dating via a comparison of a star to the isochronal sequences of representative young populations. Solar-type stars only demonstrate elevated NUV and X-ray fluxes for a short time \citep[e.g.][]{PizzolatoStellar2003}, so activity measures are unlikely to be informative for earlier type stars at all but the youngest ages. However, \citet{ShkolnikHAZMAT2014} found that the $NUV$ flux of early M dwarfs remains saturated to ages $\tau \la 300$ Myr, before then steeply declining to field levels. Thus, we expect elevated NUV activity for late-type stars in Psc-Eri (including TOI 451 B; $T_{eff} \sim 3500$ K), but not early type stars (including TOI 451, $T_{eff} \sim 5500$ K).

We consider the UV/NIR flux ratio $F_{NUV}/F_J$ as a function of spectral type, as suggested by \citet{ShkolnikSearching2011}. A similar method was also suggested by \cite{FindeisenUltravioletselected2010} based on $m_{NUV}-m_J$ and $J-K$ colors, while \citet{RodriguezNew2011} used $m_{NUV}-m_V$ and $m_{nuv}-m_J$. 

TOI 451 and TOI 451 B were detected by \galex during short NUV exposures as part of the All-Sky Imaging Survey (AIS). \citep{BianchiRevised2017} reported $m_{nuv,p} = 16.77 \pm 0.02$ mag and $m_{nuv,s} = 21.68 \pm 0.31$ mag for TOI 451 and TOI 451 B, respectively. Both stars also are clearly present in longer NUV exposures from the Medium-depth Imaging Survey (MIS), but only TOI 451's brightness was reported in the MIS catalog ($m_{nuv,p} = 16.674 \pm 0.006$ mag; \citealt{BianchiRevised2017}). Since the AIS magnitude for TOI 451 B was only measured with a significance of $3.5 \sigma$, we determined a more precise magnitude for TOI 451 B using data from the MIS. We downloaded the MIS image of the TOI 451 system from the MAST GALEX archive and performed aperture photometry, using the recommended aperture sizes and aperture corrections from \cite{MorrisseyCalibration2007}. For TOI 451, we recovered a brightness similar to the MIS catalog value ($m_{nuv,p} = 16.637 \pm 0.006$ mag). We measured the secondary's brightness to be $m_{nuv,s} = 21.33 \pm 0.06$ mag, which is consistent with the AIS value within $\sim 1 \sigma$. In our analysis, we adopt the MIS catalog value for the primary and our MIS aperture photometry value for the secondary. We also use $J$ band photometry from 2MASS \citep{CutriVizieR2003} and $B_p$ and $R_p$ photometry from Gaia DR2 \citep{GaiaCollaborationGaia2018}.

In Figure~\ref{fig:galex}, we show isochronal sequences for the other members of Psc-Eri using the membership catalog from \citet{CurtisTESS2019} and GALEX photometry from the AIS catalog \citep{BianchiRevised2017}. Since this sample did not extend to later spectral types, we supplement the Psc-Eri sequence with the similarly-aged Pleiades cluster ($\tau \sim 120$ Myr; \citealt{StaufferKeck1998}), which yields a more robust M-dwarf sequence due to its lower distance and more complete census \citep{GaiaCollaborationGaia2018}. To further extend the Pleiades sequence beyond the limit of the AIS, we also queried the GALEX catalog at MAST for photometry of Pleiades members observed in MIS fields, as well as in GO programs previously reported by \cite{BrowneGALEX2009} and \cite{FindeisenUltravioletselected2010}. Finally, to define the expected fluxes for older stars, we show the sequence of AIS catalog photometry for the Hyades cluster ($\tau \sim 680$ Myr; see \citealt{GossageAge2018} and discussion therein) based on the cluster core population identified by \cite{2019A&A...621L...2R}. 

For all clusters, we convert $B_p-R_p$ colors to spectral types using the color-spectral type relations of \cite{PecautIntrinsic2013}, as updated by E. Mamajek on 20190322\footnote{\url{http://www.pas.rochester.edu/~emamajek/EEM_dwarf_UBVIJHK_colors_Teff.txt}}. We also display the mapping of spectral type to $B_p-R_p$ on the upper axis. 
TOI 451 is indeed consistent with the sequences for other G stars in all three clusters, and hence its NUV excess is not strongly diagnostic of age. On the other hand, the early M dwarf TOI 451 B sits well above the Hyades sequence, and is instead consistent with the Pleiades sequence.

The success in applying this method to TOI 451 suggests that NUV excess--especially given the wide availability of data from GALEX--is a promising route for further growing the low-mass census of Psc--Eri. Examination of Figure \ref{fig:galex} suggests that this method is applicable from mid K to mid M dwarfs.

\end{document}